\documentclass[draftcls,12pt,onecolumn,oneside]{IEEEtran}
\usepackage{amsmath,amssymb} \interdisplaylinepenalty=2500
\usepackage{color}
\usepackage{epsfig}
\usepackage{graphicx}
\usepackage{dsfont}
\usepackage{verbatim}
\usepackage{amsthm}
\usepackage{setspace}
\usepackage{mathtools}
\usepackage{enumerate}

\usepackage{tabulary}
\usepackage{booktabs}
\usepackage{bbm}

\usepackage[table]{xcolor}
\usepackage{color, colortbl}
\usepackage{multirow,bigdelim}
\usepackage{amstext}
\usepackage{array}
\usepackage{cite}
\usepackage{arydshln}

\usepackage{graphicx}
\usepackage{subcaption}

\newtheorem{definition}{Definition}
\newtheorem{theorem}{Theorem}
\newtheorem{lemma}[theorem]{Lemma}

\newenvironment{remark}{\textit{Remark: }}{}

\newif\ifcomments
\commentstrue 

\newcommand{\calE}{\mathcal{E}}

\newcommand{\calI}{\mathcal{I}}

\newcommand{\calK}{\mathcal{K}}
\newcommand{\calL}{\mathcal{L}}

\newcommand{\calQ}{\mathcal{Q}}

\newcommand{\calS}{\mathcal{S}}
\newcommand{\calT}{\mathcal{T}}
\newcommand{\calU}{\mathcal{U}}

\newcommand{\calW}{\mathcal{W}}

\newcommand{\bfG}{\mathbf{G}}

\newcommand{\bfS}{\mathbf{S}}

\DeclarePairedDelimiterX{\norm}[1]{\lVert}{\rVert}{#1}

\newcommand{\nFile}{K}

\newcommand{\findex}{\kappa}
\newcommand{\nNode}{N}

\newcommand{\queries}{\mathcal{Q}}

\newcommand{\Fq}{\mathds{F}_{q}}

\newcommand{\Tcomp}{\overline{\mathcal{T}}}
\newcolumntype{L}{>{$}l<{$}}

\definecolor{Gray}{gray}{0.9}
\definecolor{LightCyan}{rgb}{0.88,1,1}
\definecolor{HoneydewTwo}{rgb}{0.96,0.92,0.88}

\date{}
\begin{document}

\title{Secure Private Information Retrieval from Colluding Databases with Eavesdroppers}
\author{\IEEEauthorblockN{Qiwen~Wang, and~Mikael~Skoglund} \\
    \IEEEauthorblockA{School of Electrical Engineering, KTH Royal Institute of Technology} 
    \\Email: \{qiwenw, skoglund\}@kth.se \vspace{-1em}}

\maketitle
\begin{abstract}
The problem of \emph{private information retrieval (PIR)} is to retrieve one message out of $K$ messages replicated at $N$ databases, without revealing the identity of the desired message to the databases.
We consider the problem of PIR with colluding servers and eavesdroppers, named \emph{T-EPIR}. Specifically, any $T$ out of $N$ databases may collude, that is, they may communicate their interactions with the user to guess the identity of the requested message. An eavesdropper is curious to know the database and can tap in on the incoming and outgoing transmissions of any $E$ databases.
The databases share some common randomness unknown to the eavesdropper and the user, and use the common randomness to generate the answers, such that the eavesdropper can learn no information about the $K$ messages.
Define $R^{*}$ as the optimal ratio of the number of the desired message information bits to the number of total downloaded bits, and $\rho^{*}$ to be the optimal ratio of the information bits of the shared common randomness to the information bits of the desired file.
In our previous work~\cite{wang2017secure}, we found that when $E \geq T$, the optimal ratio that can be achieved (hence is the capacity) equals $1-\frac{E}{N}$.
In this work, we focus on the case when $E \leq T$.
We derive an outer bound (converse bound) that $R^{*} \leq \left( 1- \frac{T}{N} \right) \frac{ 1- \frac{E}{N} \cdot \left(\frac{T}{N} \right)^{K-1}  }{1-\left(\frac{T}{N}\right)^K }$.
We also obtain a lower bound (converse bound) of $\rho^{*} \geq \frac{ \frac{E}{N} \left( 1-\left(\frac{T}{N}\right)^K \right)} { \left( 1- \frac{T}{N} \right) \left( 1- \frac{E}{N} \cdot \left(\frac{T}{N} \right)^{K-1}  \right) }$.
For the achievability, we propose a scheme which achieves the rate (inner bound) $R = \frac{1-\frac{T}{N}}{1 - (\frac{T}{N})^K} - \frac{E}{KN}$.
The amount of shared common randomness used in the achievable scheme is $\frac{ \frac{E}{N} \left( 1 - (\frac{T}{N})^K \right)}{ 1 - \frac{T}{N} -\frac{E}{KN} \left( 1 - (\frac{T}{N})^K \right) }$ times the file size.
The gap between the derived inner and outer bounds vanishes as the number of messages $K$ tends to infinity.
\end{abstract}

\section{Introduction}
In the situation where a user wants to retrieve a file (message) from a remotely stored database, the nature of the data might be privacy-sensitive, for example medical records, stock prices {\it etc.},  such that the user does not want to reveal the identity of the data retrieved. This is known as the problem of private information retrieval (PIR). In some cases, the privacy of the database needs also to be preserved. For example, if a user wants to retrieve his/her medical data from a database, it is hoped that the user obtains no information about other users' medical records. This is known as the problem of symmetric private information retrieval (SPIR).

The problem of PIR and SPIR was firstly studied in the computer science literature.
In~\cite{chor1995private,chor1998private}, it is shown that if the messages are stored at a single database, the only possible scheme for the user is to download all the messages to guarantee information-theoretic privacy, which is inefficient in practice. It is further shown that the communication cost can be reduced in sublinear scale by replicating the database at multiple non-colluding servers \cite{chor1998private}.
To further protect the privacy of the database such that the user obtains no more information regarding the other messages besides the requested message, the problem of SPIR is introduced~\cite{gertner1998protecting}. 
In~\cite{chor1995private,chor1998private,gertner1998protecting}, the collection of messages stored at each database is modeled as a bit string, and the user wishes to retrieve a single bit.
In these works, the communication cost is measured as the sum of the transmission at the querying phase from user to servers and at the downloading phase from servers to user.

When the message size is significantly large and the target is to minimize the communication cost of only the downloading phase, the metric of the downloading cost is defined as the number of bits downloaded per bit of the retrieved message, and the reciprocal of which is named the \emph{PIR capacity}. 
A series of recent works derive information-theoretic limits of various versions of the PIR problem
\cite{sun2017capacity,sun2016colluding,sun2016SPIR,banawan2016capacity,banawan2017multi,banawan2017capacity,wang2016symmetric} {\it etc}. 
The leading work in the area is by Sun and Jafar\cite{sun2017capacity}, where the authors find the capacity of the PIR problem with replicated databases.
In subsequent works by Sun and Jafar~\cite{sun2016colluding,sun2016SPIR}, the PIR capacity with duplicated databases and colluding servers, and the SPIR capacity with duplicated (non-colluding) databases are derived.
In~\cite{banawan2016capacity,banawan2017multi,banawan2017capacity}, Banawan and Ulukus find the capacity of the PIR problem with coded databases, multi-message PIR with replicated databases, and the PIR problem with colluding and Byzantine databases.
In our previous works~\cite{wang2016symmetric,wang2017linear,wang2017secure}, we derive the capacity of the SPIR problem with coded databases, linear SPIR with colluding and coded databases, and the SPIR problem with Byzantine adversaries and eavesdroppers.


Another series of works focus more on the coding structure of the storage system, and study schemes and information limits for various PIR problems with coded databases~\cite{shah2014one,fazeli2015pir,chan2015private,tajeddine2016private,freij2016private}.
In~\cite{shah2014one}, PIR is achieved by downloading one extra bit other than the desired file, given that the number of storage nodes grows with file size, which can be impractical in some storage systems.
In~\cite{fazeli2015pir}, storage overhead can be reduced by increasing the number of storage nodes.
In~\cite{chan2015private}, tradeoff between storage cost and downloading cost is analyzed. Subsequently in~\cite{tajeddine2016private}, explicit schemes which match the tradeoff in~\cite{chan2015private} are presented.
It is worth noting that in~\cite{banawan2016capacity}, the capacity of PIR for coded database is settled, which improves the results in~\cite{chan2015private,tajeddine2016private}.
Recently in~\cite{freij2016private}, the authors present a framework for PIR from coded databases with colluding servers.

In our previous work~\cite{wang2017secure}, we studied the problem of SPIR from replicated databases with colluding databases and eavesdroppers, named \emph{T-ESPIR}. 
Briefly speaking, a user wants to retrieve one file out of $K$ files that are replicatively stored at $N$ databases. Any $T$ out of the $N$ servers may collude, that is, they may share their communication with the user to infer the identity of the requested file. 
A passive eavesdropper is curious to know the database and can tap in on the incoming and outgoing transmissions of any $E$ servers.
In the problem of T-ESPIR, it is required that the user learns no information about the database other than the requested file.
In~\cite{wang2017secure}, we show that the information-theoretical capacity of the T-ESPIR problem is $1-\frac{\max(T,E)}{N}$, if the databases share common randomness with amount at least $\frac{\max(T,E)}{N-\max(T,E)}$ times the file size. 
In Section VI.B in~\cite{wang2017secure}, we discussed that if database-privacy is not required, {\it i.e.} the user can learn information about the other files, and when $E \geq T$, the capacity of the T-EPIR problem is $1-\frac{E}{N}$. 

In this work, we continue the study of the T-EPIR problem when $E \leq T$. 
We derive an outer bound (converse bound) that $R^{*} \leq \left( 1- \frac{T}{N} \right) \frac{ 1- \frac{E}{N} \cdot \left(\frac{T}{N} \right)^{K-1}  }{1-\left(\frac{T}{N}\right)^K }$.
We also obtain a lower bound (converse bound) of $\rho^{*} \geq \frac{ \frac{E}{N} \left( 1-\left(\frac{T}{N}\right)^K \right)} { \left( 1- \frac{T}{N} \right) \left( 1- \frac{E}{N} \cdot \left(\frac{T}{N} \right)^{K-1}  \right) }$.
For the achievability, we propose a scheme which achieves the rate (inner bound) $R = \frac{1-\frac{T}{N}}{1 - (\frac{T}{N})^K} - \frac{E}{KN}$.
The amount of shared common randomness used in the achievable scheme is $\frac{ \frac{E}{N} \left( 1 - (\frac{T}{N})^K \right)}{ 1 - \frac{T}{N} -\frac{E}{KN} \left( 1 - (\frac{T}{N})^K \right) }$ times the file size.
The capacity of T-ESPIR when $E < T$ remains an open problem. 
In Section~\ref{sec:main}, we discuss four special cases in which the capacity is known or can be easily derived, and reveal that our outer bound is tight for the four special cases. On the other hand, the inner bound is tight for three cases but one, namely, when $E=T$, the derived inner bound does not match with the capacity at this point. For illustration, we plot the results in Figure~\ref{fig:RvsT} and Figure~\ref{fig:RvsE} for some chosen parameters. It can be observed from the figures that the gap between inner and outer bounds decays and vanishes as $K$ tends to infinity.

\section{Model}
\subsection{Notation}
Let $[m:n]$ denote the set $\{m, m+1, \dots, n\}$ for $m \leq n$. To simplify the notation, denote the set of random variables $\{ X_m, X_{m+1}, \dots, X_n\}$ by $X_{[m:n]}$.
For an index set $\calI = \{ i_1, i_2, \dots, i_n \}$, denote the set of variables with the index set $\{ X_i: i \in \calI \}$ by $X_{\calI}$.
For a matrix $\bfS$, let $\bfS[:, \calI]$ denote the submatrix of $\bfS$ comprised of the columns corresponding to the index set $\calI$. 
The transpose of matrix $\bfG$ is denoted by $\bfG^{\textrm{T}}$.
Let $\sim$ denote the statistical equivalence between random variables, that is, if $X \sim Y$, then $X$ and $Y$ are identically distributed.

\subsection{Problem Description}
\noindent{\bf Replicated databases:}
A collection of $K$ independent messages (files), denoted by $W_1, \dots, W_{K}$, are replicatively stored at $N$ databases (nodes). Each message consists $L$ information bits. Therefore, for any $k \in [1:\nFile]$, 
\begin{equation}
H(W_k)=L  \quad ; \quad H(W_1, \dots, W_{K}) = KL . \nonumber
\end{equation}


\noindent{\bf User queries:}
A user wants to retrieve a message $W_{\findex}$ with index $\findex$ from the database, where the desired message index $\findex$ follows some prior distribution among $[1:K]$. 
Let $\calU$ denote a random variable privately generated by the user, which represents the randomness of the query scheme followed by the user. The random variable $\calU$ is generated independently of the messages and the desired file index. Let the realization of the file index $\findex$ be $k$, based on the realization of the desired file index $k$ and the realization of $\calU$, the user generates and sends queries to all nodes, where the query received by node-$n$ is denoted by $Q_{n}^{[k]}$. Let $\queries = [Q_{n}^{[k]}]_{n \in [1:\nNode], k \in [1:\nFile]}$ denote the complete query scheme, namely, the collection of all queries under all cases of desired message index.
We have that $H(\queries | \calU) = 0$.

\noindent{\bf Common randomness:}
Let random variable $S$ denote the common randomness shared by all databases, the realization of which is known to all the databases but unavailable to the user and the eavesdropper. 
The common randomness is utilized to protect the system-privacy~\eqref{eqn:privacy_E} below, that is, to prevent the eavesdropper from learning the messages.

\noindent{\bf Database answers:}
The databases generate answers according to the agreed scheme with the user based on the received query $Q_n^{[k]}$, the stored messages $W_{[1:K]}$, and the common randomness $S$. The answer generated and sent to the user by node $n$ is denoted by $A_{n}^{[k]}$.

\noindent{\bf Eavesdropper:}
A {\it passive eavesdropper} can tap in on the incoming and outgoing transmissions of $E$ nodes in the system. The eavesdropper is ``nice but curious," in the sense that the goal of the eavesdropper is to obtain some information about the database, without corrupting any transmission. The user has no knowledge of the identity of the nodes tapped on by the eavesdropper.


\noindent{\bf T-EPIR:}
Based on the received answers $A_{[1:N]}^{[k]}$ and the query scheme $\queries$, the user shall be able to decode the requested message $W_{k}$ with zero error. 
Any set of $T$ databases may collude to guess the requested message index, by communicating their interactions with the user.
Two privacy constraints must be satisfied:
\begin{itemize}
\item \emph{User-privacy:} any $T$ colluding databases shall not be able to obtain any information regarding the identity of the requested message, {\it i.e.,}
	\begin{equation}
		I(\findex ; Q_{\calT}^{[\findex]}, A_{\calT}^{[\findex]}, W_{[1:K]}, S ) = 0,  \forall  \calT \subset [1:\nNode], |\calT|=T. \label{eqn:user_privacy}
	\end{equation}

\item \emph{System-privacy:} For any set of databases $\calE$ with size at most $E$, and for any $k \in [1:K]$:
\begin{equation}
	I( W_{[1:K]} ; Q_{\calE}^{[k]}, A_{\calE}^{[k]}) = 0. \label{eqn:privacy_E}
\end{equation}

\end{itemize}

\begin{definition} 
The rate of a T-EPIR scheme is the number of information bits of the requested file retrieved per downloaded answer bit. By symmetry among all files, for any $k \in [1:K]$,
\begin{equation}
R_{\textrm{T-EPIR}} \triangleq \frac{H(W_{k})}{\sum_{n=1}^{N} H(A_n^{[k]})}. \nonumber
\end{equation}
The optimal rate of T-EPIR schemes is denoted by $R_{\textrm{T-EPIR}}^{*}$. The capacity $C_{\textrm{T-EPIR}}$ is the supremum of $R_{\textrm{T-EPIR}}$ over all T-EPIR schemes.
\end{definition}

\begin{definition} 
The secrecy rate is the amount of common randomness shared by the storage nodes relative to the file size, that is
\begin{equation}
\rho_{\textrm{T-EPIR}} \triangleq \frac{H(S)}{H(W_{k})}. \nonumber
\end{equation}
\end{definition}

\section{Main Result} \label{sec:main}
In this section, we summarize the main results of this paper. 

\begin{theorem}[{Capacity when $E \geq T$}]
For T-EPIR with $K$ files replicated at $N$ databases, where any $T$ nodes may collude and an eavesdropper can tap in on the communication of any $E$ nodes, when $E \geq T$, the capacity is 
\begin{equation}
C_{\textrm{T-EPIR}} = 
\begin{cases}
1-\frac{E}{N}, & \text{if } \rho_{\textrm{T-EPIR}} \geq \frac{E}{N-E}\\
0, & \text{otherwise}
\end{cases}
. \nonumber
\end{equation} 
\label{thm:EgeqT}
\end{theorem} 

\begin{remark}
For the detailed proof of Theorem~\ref{thm:EgeqT}, we refer to Section V and Section VI.B of our previous work~\cite{wang2017secure}.
\end{remark}

\begin{theorem}[Outer Bound when $E \leq T$]
For T-EPIR with $K$ files replicated at $N$ databases, where any $T$ nodes may collude and an eavesdropper can tap in on the communication of any $E$ nodes, when $E \leq T$,
\begin{equation}
 R_{\textrm{T-EPIR}}^* \leq \overline{R}_{\textrm{T-EPIR}} = \left( 1- \frac{T}{N} \right) \frac{ 1- \frac{E}{N} \cdot \left(\frac{T}{N} \right)^{K-1}  }{1-\left(\frac{T}{N}\right)^K }.
\end{equation}
The secrecy rate, {\it i.e.} the ratio of the amount of common randomness to the file size is at least $\rho_{\textrm{T-EPIR}} \geq \frac{ \frac{E}{N} \left( 1-\left(\frac{T}{N}\right)^K \right)} { \left( 1- \frac{T}{N} \right) \left( 1- \frac{E}{N} \cdot \left(\frac{T}{N} \right)^{K-1}  \right) }$.
\label{thm:outer}
\end{theorem} 
\begin{remark}
The proof of the outer bound is in Section~\ref{sec:proof_outer}.
The outer bound is tight, that is, it can be achieved and is hence the capacity of the problem for the four special cases below.
\begin{itemize}
\item Case 1 ($E=T$): From Theorem~\ref{thm:EgeqT}, the capacity is $C_{\textrm{T-EPIR}} = 1-\frac{E}{N}$ when $E = T$. The outer bound in Theorem~\ref{thm:outer} is $\overline{R}_{\textrm{T-EPIR}} = \left( 1- \frac{T}{N} \right) \frac{ 1- \frac{E}{N} \cdot \left(\frac{T}{N} \right)^{K-1}  }{1-\left(\frac{T}{N}\right)^K } = 1-\frac{T}{N} = 1-\frac{E}{N} = C_{\textrm{T-EPIR}}$ when $E=T$.

\item Case 2 ($E=0$): When there is no eavesdropper, {\it i.e.} $E=0$, the problem reduce to the TPIR problem in~\cite{sun2016colluding}, where the authors derive the capacity to be $C_{\textrm{TPIR}} = \frac{ 1-  \frac{T}{N} }{1-\left(\frac{T}{N}\right)^K } $. The outer bound in Theorem~\ref{thm:outer} is $\overline{R}_{\textrm{T-EPIR}} = \left( 1- \frac{T}{N} \right) \frac{ 1- \frac{E}{N} \cdot \left(\frac{T}{N} \right)^{K-1}  }{1-\left(\frac{T}{N}\right)^K } = \frac{ 1-  \frac{T}{N} }{1-\left(\frac{T}{N}\right)^K }  = C_{\textrm{TPIR}}$ when $E=0$.

\item Case 3 ($K \to \infty$): In our previous work~\cite{wang2017secure}, we derive the T-ESPIR capacity to be $C_{\textrm{T-ESPIR}} = 1-\frac{\max{(T,E)}}{N} = 1-\frac{T}{N}$ when $E \leq T$. 
As with all previous works for various scenarios of the PIR and SPIR problems, the PIR capacity reduces to the SPIR capacity when the number of files $K \to \infty$. 
The intuition is that, when the number of files increases, the penalty in the downloading rate to protect database-privacy for SPIR decays.
When there are asymptotically infinitely many files, the information rate the user can learn about the database from finite downloaded symbols vanishes.
When the number of files $K$ tends to infinity, the outer bound tends to $\lim_{K \to \infty} \overline{R}_{\textrm{T-EPIR}} = \lim_{K \to \infty}  \left( 1- \frac{T}{N} \right) \frac{ 1- \frac{E}{N} \cdot \left(\frac{T}{N} \right)^{K-1}  }{1-\left(\frac{T}{N}\right)^K } = 1-\frac{T}{N} = C_{\textrm{T-ESPIR}} $.

\item Case 4 ($T = N$): When all databases collude, that is $T=N$, if furthermore $E=T=N$, the capacity is $0$ because the eavesdropper receives the same information as the user. If the user can decode $W_k$, so does the eavesdropper. Hence, the problem is non-trivial only if $E$ is strictly smaller than $T$.
Suppose each file consists $L = N-E$ symbols from a large enough finite field $\Fq$, denoted by row vectors $W_k^{[1:L]}$ for $k \in [1:K]$, consider the scheme below.

The databases generate $KE$ uniformly i.i.d. symbols from $\Fq$, denoted by $K$ length-$E$ row vectors $S_k^{[1:E]}$ for $k \in [1:K]$. Let $\bfG^{E \times N}$ be the generating matrix of an $(N,E)$-MDS code. The databases operate the $(N,E)$-MDS code on the common randomness vectors to obtain $K$ length-$N$ vectors $\bar{S}_k^{[1:N]} = S_k^{[1:E]} \bfG^{E \times N}$ for $k \in [1:K]$, such that any $E$ symbols from $\bar{S}_k^{[1:N]} $ are uniformly identically distributed over $\Fq$. For each $k $, let $A_k^{[1:N]} = [\mathbf{0}^{[1\times E]} W_k^{[1:L]} ] + \bar{S}_k^{[1:N]} $ where $\mathbf{0}^{[1\times E]} $ is a length-$E$ zero vector, the user downloads $A_k^n$ from database $n$ for each file index $k$.
It can be checked that the user can decode $W_k$ (in fact the user can decode all files), and both user-privacy and system-privacy are guaranteed. The rate achieved by the scheme is $\frac{N-E}{NK}$.

The outer bound in Theorem~\ref{thm:outer} is $\overline{R}_{\textrm{T-EPIR}} = \left( 1- \frac{T}{N} \right) \frac{ 1- \frac{E}{N} \cdot \left(\frac{T}{N} \right)^{K-1}  }{1-\left(\frac{T}{N}\right)^K } = \frac{1- \frac{E}{N} \cdot \left(\frac{T}{N} \right)^{K-1} }{1+\frac{T}{N}+ \dots + \left( \frac{T}{N} \right)^{K-1}   } = \frac{1-\frac{E}{N}}{K} = \frac{N-E}{NK}$ when $T=N$, which is achieved by the scheme above.
\end{itemize}
\end{remark}

\begin{theorem}[Inner Bound when $E \leq T$]
For T-EPIR with $K$ files replicated at $N$ databases, where any $T$ nodes may collude and an eavesdropper can tap in on the communication of any $E$ nodes, when $E \leq T$,
\begin{equation}
 R_{\textrm{T-EPIR}}^* \geq \underline{R}_{\textrm{T-EPIR}} = \frac{1-\frac{T}{N}}{1 - (\frac{T}{N})^K} - \frac{E}{KN}
\end{equation}
\label{thm:inner}
\end{theorem} 
\begin{remark}
The inner bound is achieved by the scheme described in Section~\ref{sec:proof_inner}.
We discuss below the rate achieved by our scheme for the four special cases discussed above in which the outer bound in Theorem~\ref{thm:outer} is tight.
\begin{itemize}
\item Case 1 ($E=T$): The capacity of T-EPIR is $C_{\textrm{T-EPIR}} = 1-\frac{E}{N} = 1-\frac{T}{N}$ when $E = T$, that is, the rate of $1-\frac{T}{N}$ can be achieved by the scheme in our previous work~\cite{wang2017secure}. When $E = T$, the rate achieved by the scheme in Section~\ref{sec:proof_inner} is $\underline{R}_{\textrm{T-EPIR}} = \frac{1-\frac{T}{N}}{1 - (\frac{T}{N})^K} - \frac{E}{KN} = \frac{1-\frac{T}{N}}{1 - (\frac{T}{N})^K} - \frac{T}{KN} = \frac{1-\frac{T}{N}}{1 - (\frac{T}{N})^K} \cdot \Big( 1 - \frac{1}{K}\big( \frac{T}{N} + ( \frac{T}{N} )^2 + \dots + ( \frac{T}{N} )^K  \big)    \Big)$, which is strictly smaller than $1-\frac{T}{N}$ when $T \neq N$. Therefore, our scheme in Section~\ref{sec:proof_inner} is not optimal when $E = T$. In other words, the inner bound in Theorem~\ref{thm:inner} is not tight for the case $E=T$.

\item Case 2 ($E=0$): When there is no eavesdropper hence $E=0$, the rate achieved is $\underline{R}_{\textrm{T-EPIR}} = \frac{1-\frac{T}{N}}{1 - (\frac{T}{N})^K} - \frac{E}{KN} = \frac{1-\frac{T}{N}}{1 - (\frac{T}{N})^K}$, which matches with the TPIR capacity derived in~\cite{sun2016colluding}, hence is optimal.

\item Case 3 ($K \to \infty$): When the number of files $K$ tends to infinity, $\lim_{K \to \infty} \underline{R}_{\textrm{T-EPIR}} = \lim_{K \to \infty} \Big(  \frac{1-\frac{T}{N}}{1 - (\frac{T}{N})^K} - \frac{E}{KN} \Big) = 1-\frac{T}{N}$. Hence, the inner bound tends to the T-ESPIR capacity as $K \to \infty$.

\item Case 4 ($T = N$): When all databases collude hence $T=N$, the rate achieved by the scheme in this work is $\underline{R}_{\textrm{T-EPIR}} = \frac{1-\frac{T}{N}}{1 - (\frac{T}{N})^K} - \frac{E}{KN} = \frac{1}{K} - \frac{E}{KN} = \frac{N-E}{KN}$, which matches the outer bound when $T=N$, hence is optimal.

\end{itemize}
\end{remark}
%
%
%

\begin{figure}
    \centering
    \begin{subfigure}[t]{0.5\textwidth}
   		 \includegraphics[width=\textwidth]{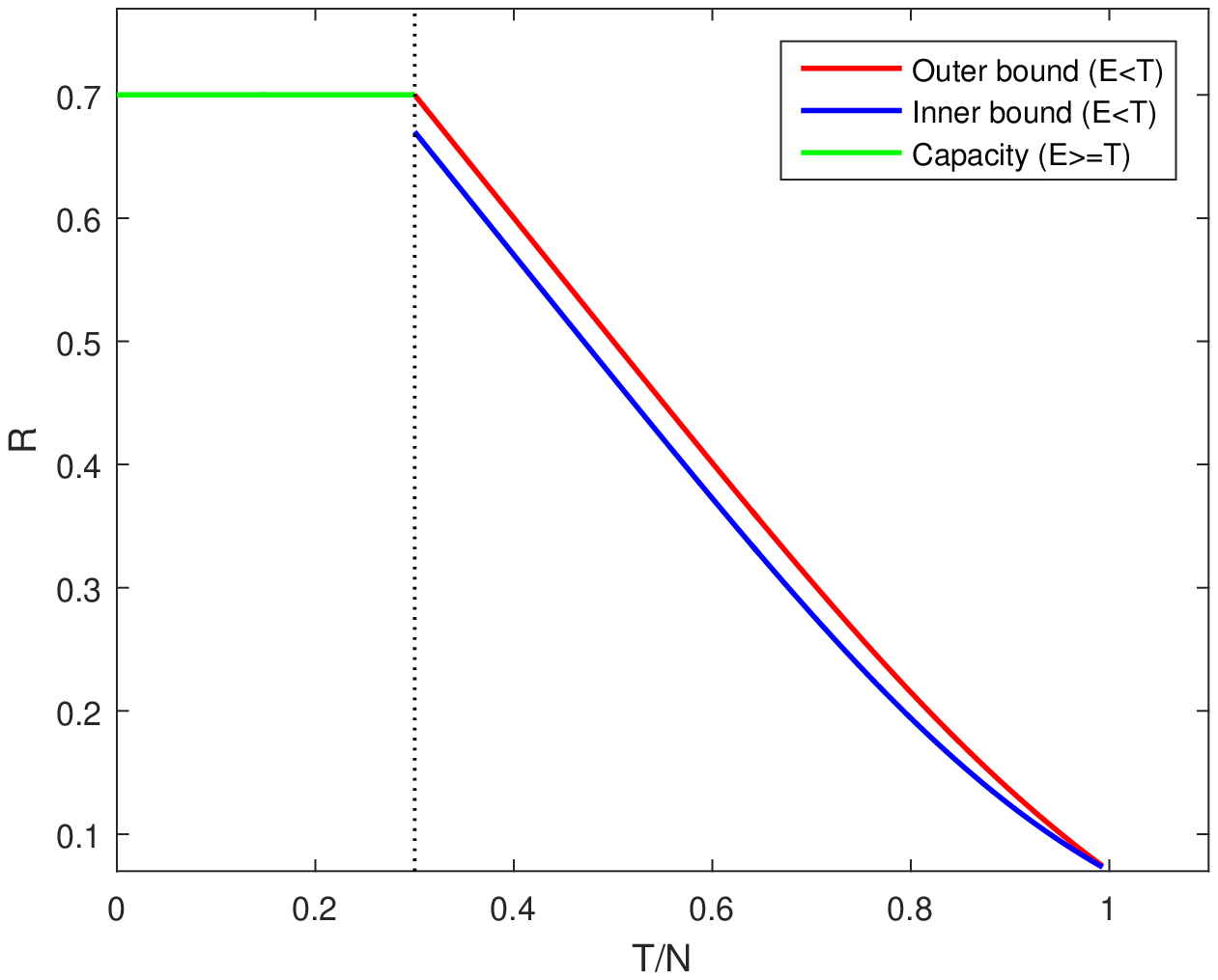}
    	 \caption{$N=10, E=3, K=10$}
    	 \label{fig:RvsT1}
    \end{subfigure}
    ~ 
    \begin{subfigure}[t]{0.47\textwidth}
        \includegraphics[width=\textwidth]{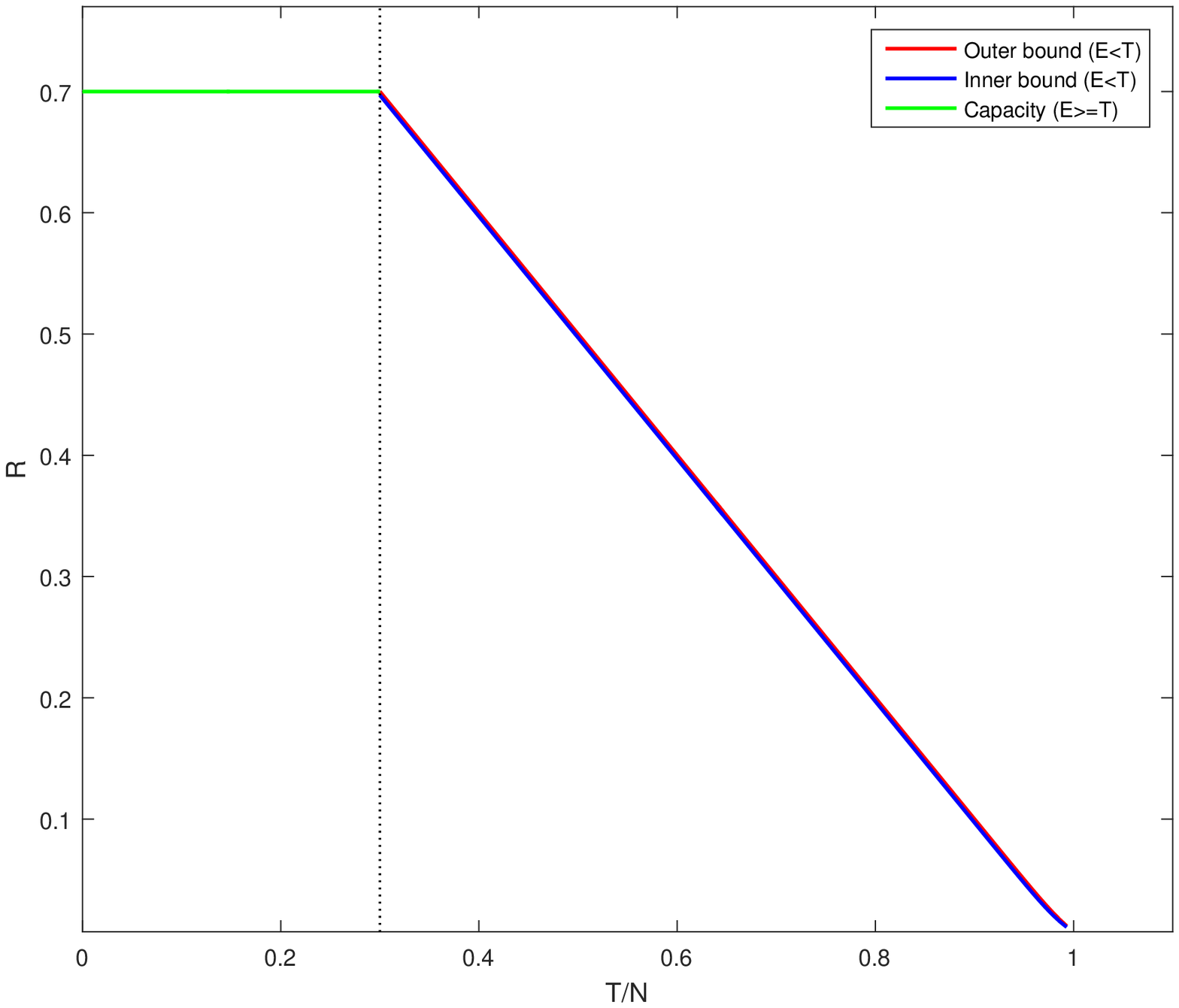}
    	 \caption{$N=10, E=3, K=100$}
    	 \label{fig:RvsT2}
    \end{subfigure}
    \caption{Plot of the bounds as functions of $\frac{T}{N}$.} \label{fig:RvsT}
\end{figure}
\begin{figure}
    \centering
    \begin{subfigure}[t]{0.48\textwidth}
   		 \includegraphics[width=\textwidth]{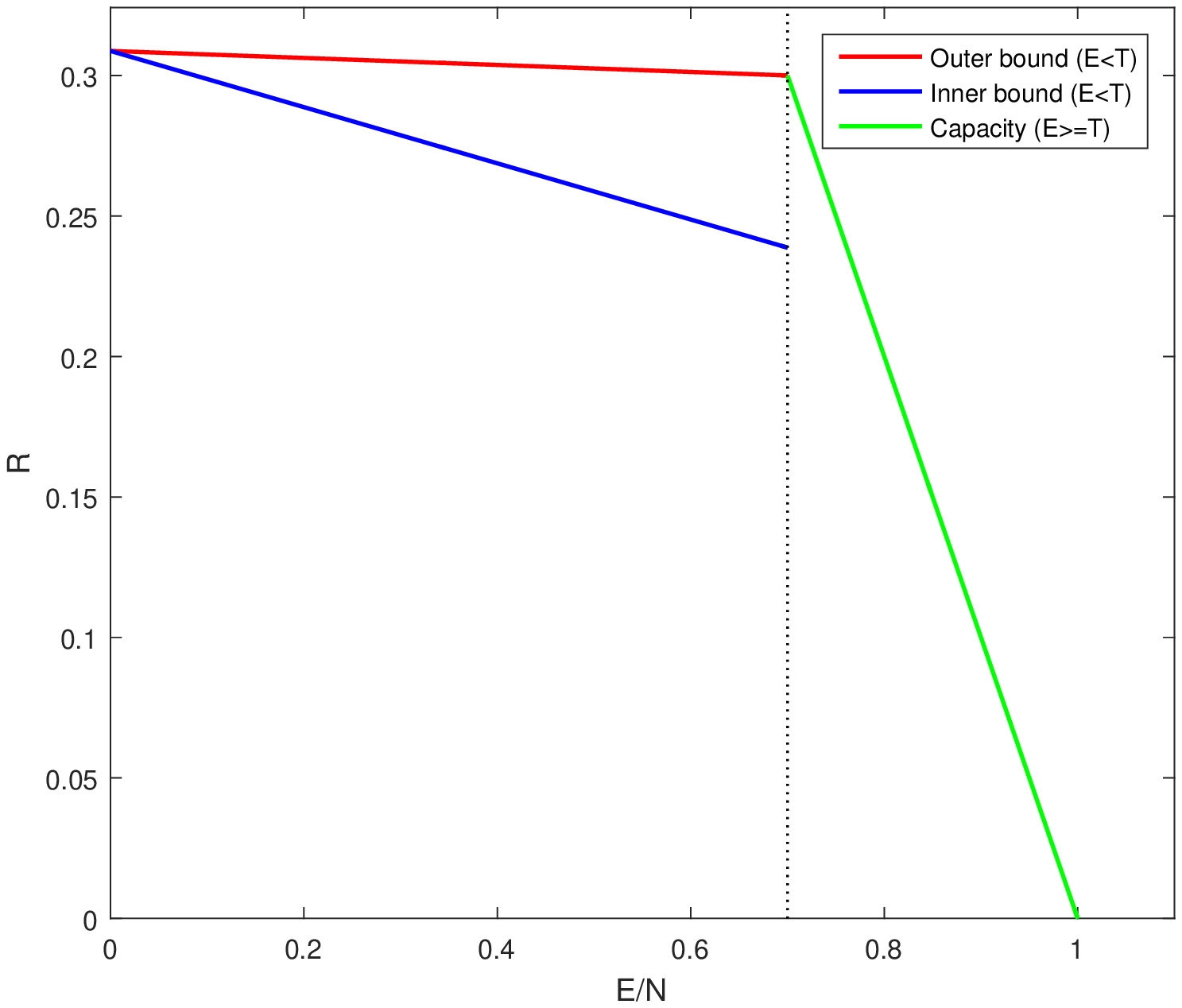}
    	 \caption{$N=10, T=7, K=10$}
    	 \label{fig:RvsE1}
    \end{subfigure}
    ~ 
    \begin{subfigure}[t]{0.48\textwidth}
        \includegraphics[width=\textwidth]{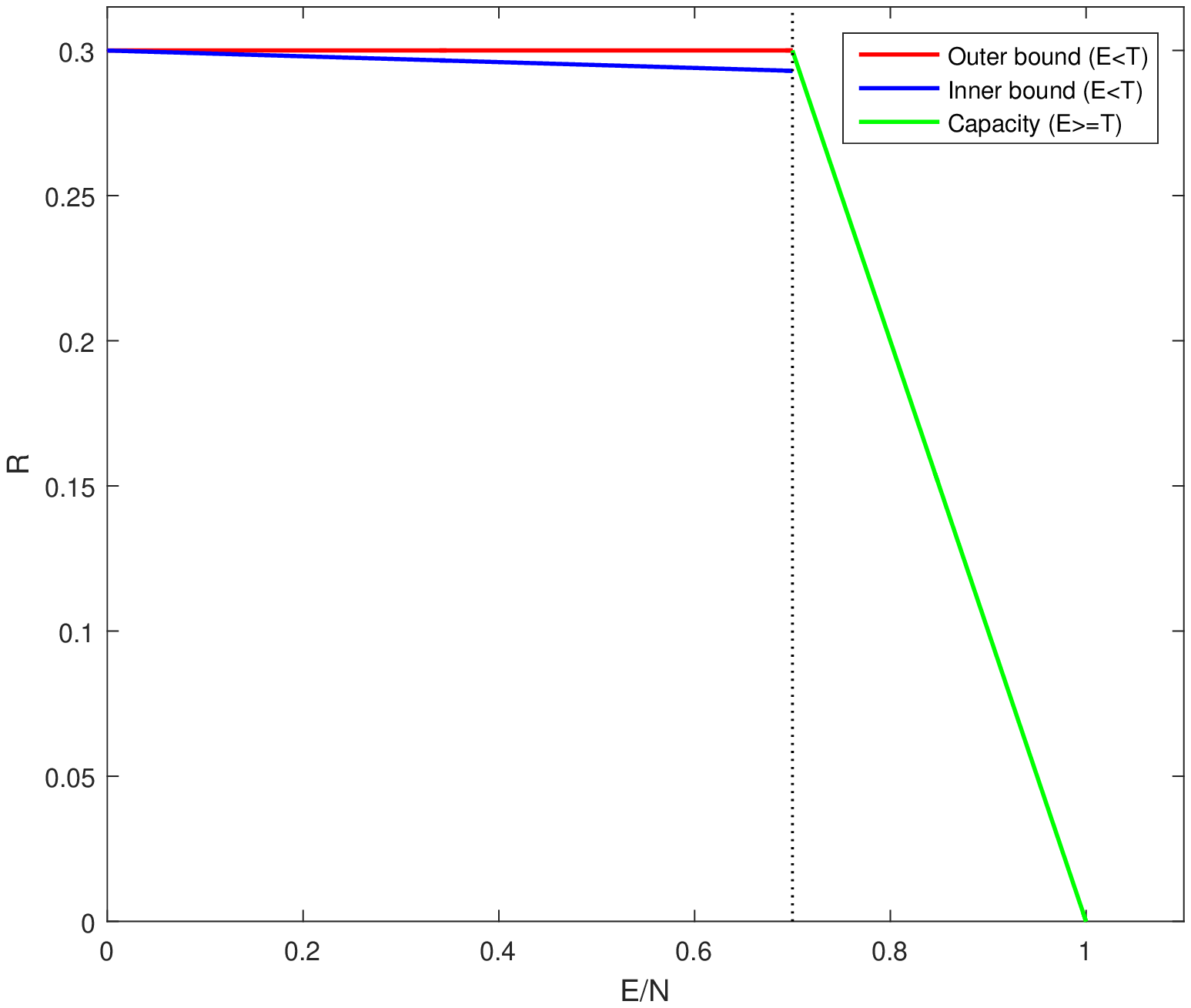}
    	 \caption{$N=10, T=7, K=100$}
    	 \label{fig:RvsE2}
    \end{subfigure}
    \caption{Plot of the bounds as functions of $\frac{E}{N}$.} \label{fig:RvsE}
\end{figure}

In Figure~\ref{fig:RvsT} and Figure~\ref{fig:RvsE}, the results of Theorems~\ref{thm:EgeqT}-~\ref{thm:inner} are plotted for several sets of parameters. It can be observed from the figures that when the number of messages $K$ increases, the gap between the inner and outer bounds decays and vanishes as $K \to \infty$.

\section{Outer bound when $E \leq T$} \label{sec:proof_outer}
In this section, we derive the outer bound presented in Theorem~\ref{thm:outer} for the PIR problem with $T$-colluding databases and $E$-eavesdropped databases when $E \leq T$. We start from the case when $K=1$ and $K=2$, then generalize to the case of arbitrary $K$ in Section~\ref{sec:outer_general}.

\subsection{$K=1$ Message} \label{sec:converseK1}
For any set of nodes $\calE \subset [1:N]$ with $|\calE| = E$,
\begin{align}
L = H(W_1) 
& = H(W_1 | \queries) - H(W_1 | A_{[1:N]}^{[1]}, \queries)\\
& = I(W_1 ; A_{[1:N]}^{[1]} | \queries) \\
& = H(A_{[1:N]}^{[1]} | \queries) - H(A_{[1:N]}^{[1]} | W_1, \queries) \\
& \leq H(A_{[1:N]}^{[1]} | \queries) - H(A_{\calE}^{[1]} | W_1, \queries) \\
& = H(A_{[1:N]}^{[1]} | \queries) - H(A_{\calE}^{[1]} | \queries) , \label{eqn:converseK1_1}
\end{align}
where~\eqref{eqn:converseK1_1} follows from system-privacy~\eqref{eqn:privacy_E}.
Averaging over all $\calE$ with size $E$ from $[1:N]$, we have that
\begin{equation}
L \leq H(A_{[1:N]}^{[1]} | \queries) - \frac{1}{{N \choose E}}\sum_{\substack{ \calE \subset [1:N] \\ |\calE|=E}} H(A_{\calE}^{[1]} | \queries). \label{eqn:converseK1_average}
\end{equation}
By Han's inequality~\cite{cover2012elements}, 
\begin{equation}
\frac{1}{{N \choose E}}\sum_{\substack{ \calE \subset [1:N] \\ |\calE|=E}} H(A_{\calE}^{[1]} | \queries) \geq \frac{E}{N} H(A_{[1:N]}^{[1]} | \queries). \label{eqn:converseK1_Han}
\end{equation}
Therefore, $L \leq \left(1 - \frac{E}{N}\right) H(A_{[1:N]}^{[1]} | \queries)$and hence $R = \frac{L}{\sum_{n=1}^{N} H(A_n^{[1]})} \leq \frac{L}{H(A_{[1:N]}^{[1]} | \queries)} \leq 1 - \frac{E}{N}$.

\subsection{$K=2$ Messages} \label{sec:converseK2}
For any set of nodes $\calT \subset [1:N]$ with $|\calT|=T$, because of user-privacy, we can ignore the requested file index of $A_{\calT}$,
\begin{align}
L = H(W_1)
& = H(W_1) - H(W_1 | A_{[1:N]}^{[1]}, \queries) \\
& = H( A_{[1:N]}^{[1]} | \queries) - H(A_{[1:N]}^{[1]}| W_1, \queries) \\
& \leq H( A_{[1:N]}^{[1]} | \queries) - H(A_{\calT}| W_1, \queries) \\
& \leq  H( A_{[1:N]}^{[1]} | \queries) - \frac{T}{N} H(A_{[1:N]}^{[2]}| W_1, \queries), \label{eqn:converseK2_1}
\end{align}
where the last step~\eqref{eqn:converseK2_1} is obtained by averaging over all $\calT$ with size $T$ and applying Han's inequality, similarly as~\eqref{eqn:converseK1_average} and~\eqref{eqn:converseK1_Han} in Section~\ref{sec:converseK1}.
hence, we have that
\begin{equation}
	H(A_{[1:N]}^{[2]}| W_1, \queries) \leq \frac{N}{T} \left(  H( A_{[1:N]}^{[1]} | \queries)- L  \right). \label{eqn:converseK2_condition}
\end{equation}

For any set of nodes $\calE \subset [1:N]$ with $|\calE| = E$, and any set of nodes $\calT \subset [1:N]$ with $|\calT|=T$,
\begin{align}
2L 
& = H(W_1, W_2) \\
& = H(W_1, W_2 | \queries) -   H(W_1, W_2 | A_{[1:N]}^{[1]}, A_{[1:N]}^{[2]}, \queries) \\
& = I(W_1, W_2 ; A_{[1:N]}^{[1]}, A_{[1:N]}^{[2]} | \queries) \\
& = H(A_{[1:N]}^{[1]}, A_{[1:N]}^{[2]} | \queries) - H(A_{[1:N]}^{[1]}, A_{[1:N]}^{[2]} | W_1, W_2, \queries) \\
& \leq H(A_{[1:N]}^{[1]}, A_{[1:N]}^{[2]} | \queries) - H(A_{\calE} | W_1, W_2, \queries) \label{eqn:converseK2_2} \\ 
& = H(A_{[1:N]}^{[1]}, A_{[1:N]}^{[2]} | \queries) - H(A_{\calE} | \queries) \label{eqn:converseK2_3} \\
& = H(A_{[1:N]}^{[1]} | \queries) + H(A_{[1:N]}^{[2]} | A_{[1:N]}^{[1]} , \queries)- H(A_{\calE} | \queries) \\
& = H(A_{[1:N]}^{[1]} | \queries) + H(A_{[1:N]}^{[2]} | A_{[1:N]}^{[1]} , W_1, \queries)- H(A_{\calE} | \queries) \label{eqn:converseK2_4} \\
& \leq H(A_{[1:N]}^{[1]} | \queries) + H(A_{[1:N]}^{[2]} | A_{\calT} , W_1, \queries)- H(A_{\calE} | \queries) \\
& =  H(A_{[1:N]}^{[1]} | \queries) + H(A_{[1:N]}^{[2]} |  W_1, \queries) - H(A_{\calT} | W_1, \queries) - H(A_{\calE} | \queries) \\
& \leq H(A_{[1:N]}^{[1]} | \queries) + \left(1-\frac{T}{N} \right) H(A_{[1:N]}^{[2]} |  W_1, \queries)  - H(A_{\calE} | \queries) \label{eqn:converseK2_5} \\
& \leq  H(A_{[1:N]}^{[1]} | \queries) + \left(1-\frac{T}{N} \right) \frac{N}{T} \left(  H( A_{[1:N]}^{[1]} | \queries)- L  \right)  - H(A_{\calE} | \queries) \label{eqn:converseK2_6} \\
& \leq  H(A_{[1:N]}^{[1]} | \queries) + \left(1-\frac{T}{N} \right) \frac{N}{T} \left(  H( A_{[1:N]}^{[1]} | \queries)- L  \right)  - \frac{E}{N} H( A_{[1:N]}^{[1]} | \queries) \label{eqn:converseK2_7}  \\
& = \left( \frac{N}{T} - \frac{E}{N} \right) H( A_{[1:N]}^{[1]} | \queries) - \left( \frac{N}{T} -1 \right) L,
\end{align}
where in~\eqref{eqn:converseK2_2} we can omit the message index because $\calE$ is a set with size $E \leq T$ .~\eqref{eqn:converseK2_3} follows from system-privacy~\eqref{eqn:privacy_E}. ~\eqref{eqn:converseK2_4} is due to the fact that the user can decode $W_1$ from $A_{[1:N]}^{[1]}$ and $\queries$. ~\eqref{eqn:converseK2_5} is obtained by averaging over all $\calT$ with size $T$ and applying Han's inequality.~\eqref{eqn:converseK2_6} follows from~\eqref{eqn:converseK2_condition}.~\eqref{eqn:converseK2_7} is obtained by averaging over all $\calE$ with size $E$ and applying Han's inequality.

Therefore, we have that $\left( \frac{N}{T} - \frac{E}{N} \right) H( A_{[1:N]}^{[1]} | \queries) \geq \left( \frac{N}{T} +1 \right) L$ and 
\begin{equation}
	R = \frac{L}{\sum_{n=1}^{N} H(A_n^{[1]})} \leq \frac{L}{H(A_{[1:N]}^{[1]} | \queries)} \leq \frac{1- \frac{E}{N} \cdot \frac{T}{N}}{1+\frac{T}{N}}.
\end{equation}

\subsection{$K \geq 3$ Messages} \label{sec:outer_general}
For any set of nodes $\calT \subset [1:N]$ with $|\calT|=T$, and its compliment set $\Tcomp = [1:N] \setminus \calT$, and for any $k \in [2:K]$,
\begin{align}
& \quad H(A_{\Tcomp}^{[k]} |A_{\calT}, W_{[1:k-1]},\queries) \\
& = H(A_{[1:N]}^{[k]} | W_{[1:k-1]},\queries) - H(A_{\calT}| W_{[1:k-1]},\queries) \\
& \leq \left(  1 - \frac{T}{N} \right) H(A_{[1:N]}^{[k]} | W_{[1:k-1]},\queries) , \label{eqn:converseK_0}
\end{align}
where the last step follows by averaging over all $\calT$ with size $T$ and applying Han's inequality.

From $A_{[1:N]}^{[1]}, \dots, A_{[1:N]}^{[k-1]}$, the user can decode $W_{[1:k-1]}$, hence
\begin{align}
(k-1)L 
& = I(W_{[1:k-1]}; A_{[1:N]}^{[1]}, \dots, A_{[1:N]}^{[k-1]} | \queries) \\
& = H(A_{[1:N]}^{[1]}, \dots, A_{[1:N]}^{[k-1]}|\queries) - H(A_{[1:N]}^{[1]}, \dots, A_{[1:N]}^{[k-1]}|W_{[1:k-1]}, \queries) \\
& \leq H(A_{[1:N]}^{[1]}, \dots, A_{[1:N]}^{[k-1]}|\queries) - H(A_{\calT}|W_{[1:k-1]}, \queries) \label{eqn:converseK_1} \\
& \leq H(A_{[1:N]}^{[1]}, \dots, A_{[1:N]}^{[k-1]}|\queries) - \frac{T}{N} H(A_{[1:N]}^{[k]}|W_{[1:k-1]}, \queries), \label{eqn:converseK_2}
\end{align}
where in~\eqref{eqn:converseK_1}, we can omit the message index of $A_{\calT}$ because from user-privacy, the answers of any $T$ databases are independent of the message index.
Similar as above, the last step follows by averaging over all $\calT$ with size $T$ and applying Han's inequality. Because $A_{\calT}$ is independent of the message index, we can set the index to $k$ in the last step.

Therefore, from~\eqref{eqn:converseK_0} and~\eqref{eqn:converseK_2}, for any $k \in [2:K]$,
\begin{align}
& \quad H(A_{\Tcomp}^{[k]} |A_{\calT}, W_{[1:k-1]},\queries) \\
& \leq  \left(  1 - \frac{T}{N} \right)	H(A_{[1:N]}^{[k]}|W_{[1:k-1]}, \queries) \\
&  \leq \left(  1 - \frac{T}{N} \right) \frac{N}{T} \left(  H(A_{[1:N]}^{[1]}, \dots, A_{[1:N]}^{[k-1]}|\queries) - (k-1)L \right) \\
& = \left( \frac{N}{T} -1 \right) \left(  H(A_{\calT}, A_{\Tcomp}^{[1]}, \dots, A_{\Tcomp}^{[k-1]}|\queries) - (k-1)L \right) \\
& = \left( \frac{N}{T} -1 \right) \left( H(A_{[1:N]}^{[1]} | \queries) +  H(A_{\Tcomp}^{[2]}, \dots, A_{\Tcomp}^{[k-1]}| A_{\calT}, A_{\Tcomp}^{[1]},W_1, \queries) - (k-1)L \right) \label{eqn:converseK_3} \\
& \leq  \left( \frac{N}{T} -1 \right)  \left( H(A_{[1:N]}^{[1]} | \queries) +  H(A_{\Tcomp}^{[2]}, \dots, A_{\Tcomp}^{[k-1]}| A_{\calT}, W_1, \queries) - (k-1)L \right) \\
& \leq \left( \frac{N}{T} -1 \right)  \left( H(A_{[1:N]}^{[1]} | \queries) +  H(A_{\Tcomp}^{[2]} | A_{\calT}, W_1, \queries) + \dots +  H(A_{\Tcomp}^{[k-1]} | A_{\calT}, W_{[1:k-2]}, \queries)- (k-1)L \right) \label{eqn:converseK_iterate} ,
\end{align}
where~\eqref{eqn:converseK_3} holds because from $A_{\calT}, A_{\Tcomp}^{[1]}$ and $\queries$ one can decode $W_1$. The last step is obtained by repeating the chain rule and by the fact that from $A_{\calT}, A_{\Tcomp}^{[i]}$ and $\queries$ one can decode $W_i$ for $i = [2:k-2]$.

For any set of nodes $\calE \subset [1:N]$ with $|\calE| = E$,
\begin{align}
& \quad KL = H(W_{[1:K]}) \\
& = I(W_{[1:K]} ; A_{\calT}, A_{\Tcomp}^{[1]}, \dots, A_{\Tcomp}^{[K]} | \queries) \\
& = H( A_{\calT}, A_{\Tcomp}^{[1]}, \dots, A_{\Tcomp}^{[K]} | \queries) - H( A_{\calT}, A_{\Tcomp}^{[1]}, \dots, A_{\Tcomp}^{[K]} | W_{[1:K]}, \queries) \\
& \leq H( A_{\calT}, A_{\Tcomp}^{[1]}, \dots, A_{\Tcomp}^{[K]}| \queries) - H(A_{\calE} | W_{[1:K]}, \queries) \\
& = H(A_{\calT}, A_{\Tcomp}^{[1]}, \dots, A_{\Tcomp}^{[K]} | \queries) - H(A_{\calE} | \queries) \label{eqn:converseK_4} \\
& = H(A_{[1:N]}^{[1]} | \queries) + H(A_{\Tcomp}^{[2]}, \dots, A_{\Tcomp}^{[K]} | A_{[1:N]}^{[1]}, \queries) - H(A_{\calE} | \queries) \label{eqn:converseK_5} \\
& = H(A_{[1:N]}^{[1]} | \queries) + H(A_{\Tcomp}^{[2]}, \dots, A_{\Tcomp}^{[K]} | A_{[1:N]}^{[1]}, W_1, \queries) - H(A_{\calE} | \queries) \\
& \leq H(A_{[1:N]}^{[1]} | \queries) + H(A_{\Tcomp}^{[2]}, \dots, A_{\Tcomp}^{[K]} | A_{\calT}, W_1, \queries) - H(A_{\calE} | \queries) \\
& = H(A_{[1:N]}^{[1]} | \queries) + H(A_{\Tcomp}^{[2]} | A_{\calT}, W_1, \queries) + H(A_{\Tcomp}^{[3]}, \dots, A_{\Tcomp}^{[K]} | A_{\Tcomp}^{[2]}, A_{\calT}, W_1, \queries)  - H(A_{\calE} | \queries) \\
& = H(A_{[1:N]}^{[1]} | \queries) + H(A_{\Tcomp}^{[2]}| A_{\calT}, W_1, \queries) + H(A_{\Tcomp}^{[3]}, \dots, A_{\Tcomp}^{[K]}  | A_{\Tcomp}^{[2]}, A_{\calT}, W_1, W_2, \queries)  - H(A_{\calE} | \queries) \\
& \leq H(A_{[1:N]}^{[1]} | \queries) + H(A_{\Tcomp}^{[2]} | A_{\calT}, W_1, \queries) + H(A_{\Tcomp}^{[3]}, \dots, A_{\Tcomp}^{[K]}  | A_{\calT}, W_1, W_2, \queries)  - H(A_{\calE} | \queries) \\
& \leq H(A_{[1:N]}^{[1]} | \queries) + H(A_{\Tcomp}^{[2]} | A_{\calT}, W_1, \queries) + H(A_{\Tcomp}^{[3]}, | A_{\calT}, W_1, W_2, \queries)  + \dots \\
& \quad + H(A_{\Tcomp}^{[K]}, | A_{\calT}, W_{[1:K-1]}, \queries) - H(A_{\calE} | \queries) \label{eqn:converseK_6}  \\
& \leq H(A_{[1:N]}^{[1]} | \queries) + H(A_{\Tcomp}^{[2]} | A_{\calT}, W_1, \queries) + H(A_{\Tcomp}^{[3]}, | A_{\calT}, W_1, W_2, \queries)  + \dots + \left(  \frac{N}{T} -1 \right) \big( H(A_{[1:N]}^{[1]} | \queries) +  \\
& \quad H(A_{\Tcomp}^{[2]} | A_{\calT}, W_1, \queries) + \dots +  H(A_{\Tcomp}^{[K-1]} | A_{\calT}, W_{[1:K-2]}, \queries)- (K-1)L \big) - H(A_{\calE} | \queries) \label{eqn:converseK_7}  \\
& = \frac{N}{T} \big( H(A_{[1:N]}^{[1]} | \queries) + H(A_{\Tcomp}^{[2]} | A_{\calT}, W_1, \queries) + \dots +  H(A_{\Tcomp}^{[K-1]} | A_{\calT}, W_{[1:K-2]}, \queries) \big) - \\
& \quad \left( \frac{N}{T} - 1  \right) (K-1)L- H(A_{\calE} | \queries) \\
& \leq \left( \frac{N}{T} \right)^{K-1} H(A_{[1:N]}^{[1]} | \queries) - H(A_{\calE} | \queries) - \left(  1 - \frac{T}{N} \right) \Big[ \frac{N}{T} (K-1)L + \left(\frac{N}{T}\right)^2 (K-2)L \\
& \quad + \dots +\left( \frac{N}{T} \right)^{K-1} L \Big]  \label{eqn:converseK_8}  \\
& = \left( \frac{N}{T} \right)^{K-1} H(A_{[1:N]}^{[1]} | \queries) - H(A_{\calE} | \queries) - \frac{\left(\frac{N}{T}\right)^K - \frac{N}{T}}{\frac{N}{T}-1} L - (K-1)L \\
& \leq \left( \left( \frac{N}{T} \right)^{K-1} - \frac{E}{N} \right) H(A_{[1:N]}^{[1]} | \queries)  - \frac{\left(\frac{N}{T}\right)^K - \frac{N}{T}}{\frac{N}{T}-1} L - (K-1)L ,
\end{align}
where~\eqref{eqn:converseK_4} is due to system-privacy~\eqref{eqn:privacy_E}.
Steps~\eqref{eqn:converseK_5}-\eqref{eqn:converseK_6} follows by repeating the chain rule and by the fact that from $A_{\calT}, A_{\Tcomp}^{[i]}$ and $\queries$ one can decode $W_i$ for $i = [1:K-1]$.
Step~\eqref{eqn:converseK_7} follows by using inequality~\eqref{eqn:converseK_iterate} for $k=K$. 
By iteratively using inequality~\eqref{eqn:converseK_iterate} for $k=\{ K-1, K-2,\dots, 2 \}$, we obtain~\eqref{eqn:converseK_8}.
The last step follows by averaging over all $\calE$ with size $E$ and applying Han's inequality.

Therefore, $\Big( \left( \frac{N}{T} \right)^{K-1} - \frac{E}{N} \Big) H(A_{[1:N]}^{[1]} | \queries) \geq \frac{\left(\frac{N}{T}\right)^K - 1}{\frac{N}{T}-1} L $, and hence
\begin{align}
	R = \frac{L}{\sum_{n=1}^{N} H(A_n^{[1]})} \leq \frac{L}{H(A_{[1:N]}^{[1]} | \queries)} 
	& \leq \left( \frac{N}{T}-1 \right) \frac{\left( \frac{N}{T} \right)^{K-1} - \frac{E}{N} }{\left(\frac{N}{T}\right)^K - 1} \\
	& = \left( 1- \frac{T}{N} \right) \frac{ 1- \frac{E}{N} \cdot \left(\frac{T}{N} \right)^{K-1}  }{1-\left(\frac{T}{N}\right)^K } \\
	& = \overline{R}_{\textrm{T-EPIR}} .
\end{align}

To obtain a lower bound on the amount of common randomness needed to guarantee system-privacy, for any set of nodes $\calE \subset [1:N]$ with size $|\calE|=E$,
\begin{align}
0 
& = I(A_{\calE} ; W_{[1:K]} | \queries) \\
& = H(A_{\calE} | \queries) - H(A_{\calE} |W_{[1:K]} ,\queries) \\
& = H(A_{\calE} | \queries) - H(A_{\calE} |W_{[1:K]} ,\queries) + H(A_{\calE} |W_{[1:K]} ,S, \queries) \label{eqn:converseS_1} \\
& = H(A_{\calE} | \queries) - I(S; A_{\calE} |W_{[1:K]} ,\queries) \\
& = H(A_{\calE} | \queries) - H(S |W_{[1:K]} ,\queries) +H(S| A_{\calE} ,W_{[1:K]} ,\queries) \\
& \geq H(A_{\calE} | \queries) - H(S),
\end{align}
where~\eqref{eqn:converseS_1} holds because $A_{\calE}$ is a deterministic function of $W_{[1:K]}, S$ and $\queries$.
By averaging over all $\calE$ with size $E$ and applying Han's inequality,
\begin{align}
H(S) 
& \geq  \frac{1}{{N \choose E}}\sum_{\substack{ \calE \subset [1:N] \\ |\calE|=E}} H(A_{\calE} | \queries) 
   \geq \frac{E}{N} H(A_{[1:N]}^{[1]} | \queries) \\
& \geq \frac{ \frac{E}{N} \left( 1-\left(\frac{T}{N}\right)^K \right)} { \left( 1- \frac{T}{N} \right) \left( 1- \frac{E}{N} \cdot \left(\frac{T}{N} \right)^{K-1}  \right) } L.
\end{align}
Therefore, $\rho_{\textrm{T-EPIR}} = \frac{H(S)}{L} \geq \frac{ \frac{E}{N} \left( 1-\left(\frac{T}{N}\right)^K \right)} { \left( 1- \frac{T}{N} \right) \left( 1- \frac{E}{N} \cdot \left(\frac{T}{N} \right)^{K-1}  \right) }$.

\section{Inner bound when $E \leq T$} \label{sec:proof_inner}
In this section, we present an achievable scheme for the case when the eavesdropper can tap in on any $E$ databases where $E \leq T$.
The scheme is modified from the TPIR scheme in~\cite{sun2016colluding}, by downloading $K$ rounds where each round use the scheme in~\cite{sun2016colluding} with different part of the files and different part of the common randomness generated by the databases. The three principles in~\cite{sun2016colluding} still apply in our scheme.
\begin{enumerate}
\item Symmetry across databases
\item Symmetry of file indices within the queries to each database
\item Exploiting the side information of undesired files to retrieve the desired file information 
\end{enumerate}
Specifically, the new ingredient of our scheme lies in iterating the scheme in $K$ rounds to ensure each file is mixed with the common randomness in the same way, hence to fulfill principle 2.
In the following, we firstly introduce five examples. We explain in details of the examples in Section~\ref{sec:ex1} and Section~\ref{sec:ex5} about the decodability of the scheme, the guarantee of user-privacy and system-privacy, and only show the construction of the other three examples. Finally in Section~\ref{sec:general_scheme}, we show the scheme for general parameters of $N, K, T, E$.

We first reprise the following lemma from~\cite{sun2016colluding}. The lemma states that by multiplying deterministic full rank matrices on uniformly i.i.d. random matrices, the statistics of the random matrices remain unchanged. The proof can be found in~\cite{sun2016colluding}.

\begin{lemma}[~\cite{sun2016colluding}]
	Let $\bfS_1, \bfS_2, \dots, \bfS_K \in \Fq^{\alpha \times \alpha}$ be $K$ random matrices, drawn independently and uniformly from all $\alpha \times \alpha$ full-rank matrices over $\Fq$. Let $\bfG_1, \bfG_2, \dots, \bfG_K \Fq^{\beta \times \beta}$ be $K$ invertible square matrices of dimension $\beta \times \beta$ over $\Fq$ where $\beta \leq \alpha$. Let $\calI_1, \calI_2, \dots, \calI_K \in \mathbb{N}^{1 \times \beta}$ be $K$ index vectors, each containing $\beta$
 distinct indices from $[1:\alpha]$, then
 \begin{equation}
 	\left( \bfS_1[:, \calI_1] \bfG_1,  \bfS_2[:, \calI_2] \bfG_2, \dots, \bfS_K[:, \calI_K] \bfG_K  \right)
 	\sim
 	\left( \bfS_1[:, (1:\beta)], \bfS_2[:, (1:\beta)], \dots, \bfS_K[:, (1:\beta)]  \right)
 \end{equation}
 where $\bfS_i[:, \calI_i]$ denotes the $\alpha \times \beta$ matrix comprised of the columns of $\bfS_i$ with indices in $\calI_i$, and $\sim$ denotes the relation that the random variables on both sides are identically distributed.
\end{lemma}

\subsection{Example: $N = 3$ databases, $K = 2$ files, $T = 2$ colluding databases, $E = 1$ eavesdropped database} \label{sec:ex1}
Suppose each file contains $L = 13$ symbols from a sufficiently large finite field $\Fq$, $W_1 = W_1^{[1:13]}$ and $W_2 = W_2^{[1:13]}$ are represented as length-$13$ vectors over $\Fq$. W.l.o.g., assume the user wants to retrieve $W_1$.

The user downloads in two rounds, with $15$ symbols in each round as described in Table~\ref{tb:ex1} and with detailed formulation below. The databases generate $10$ uniformly random symbols, $5$ for each round, denoted as $( S_{[1:5]}^{(1)}, S_{[1:5]}^{(2)} )$. 
The scheme achieves the rate $R = 13/30$.

Let $\{ \lambda_1, \dots, \lambda_9 \}$ be $9$ distinct nonzero elements from $\Fq$. Let $\bfG_{[1:7]}^{7 \times 9}$ and $\bfG_{[8:9]}^{2 \times 9}$ be two generating matrices of MDS codes as follows,
\begin{equation}
{
\bfG_{[1:7]}^{7 \times 9} = 
\begin{bmatrix}
1 & 1 & \dots & 1\\
\lambda_1 & \lambda_2 & \dots & \lambda_9        \\
\vdots  & \vdots   & \ddots & \vdots   \\
\lambda_1^{6} & \lambda_2^{6} &  \dots & \lambda_9^{6}   
\end{bmatrix}
}
, 
\end{equation}

\begin{align}
\bfG_{[8:9]}^{2 \times 9}
& = 
\begin{bmatrix}
1 & 1 & \dots & 1\\
\lambda_1 & \lambda_2 & \dots & \lambda_9
\end{bmatrix}   \cdot diag( \lambda_1^{7}, \lambda_2^{7}, \dots, \lambda_9^{7}) \\ 
& = 
\begin{bmatrix}
\lambda_1^{7} & \lambda_2^{7} &  \dots & \lambda_9^{7}\\
\lambda_1^{8} & \lambda_2^{8} & \dots & \lambda_9^{8}         
\end{bmatrix}
. 
\end{align}

Let $\bfG = [\bfG_{[1:7]}^{7 \times 9} \;  \bfG_{[8:9]}^{2 \times 9} ]^{\textrm{T}}$, then $\bfG$ is a $9 \times 9$ invertible matrix. Similarly, let $\bfG_{[1:6]}^{6 \times 9}$ and $\bfG_{[7:9]}^{3 \times 9}$ be composed of the first six rows and the last three rows of $\bfG$ respectively.

The user privately generates matrices  $\bfS_1, \bfS_2  ,\bfS_3, \bfS_4 \in \Fq^{9 \times 9}$ uniformly and independently from all $9 \times 9$ invertible matrices over $\Fq$.

Let $\bfG_1^{6 \times 9}$ be the generating matrix of a $(9,6)$-MDS code.

\begin{table}
\begin{center}
\begin{tabular}{|c|c|c|}
\hline
DB1 & DB2 & DB3 \\
\hline
$a_1^{(r)}, a_2^{(r)}$ & $a_3^{(r)}, a_4^{(r)}$ & $a_5^{(r)}, a_6^{(r)}$ \\
$b_1^{(r)}, b_2^{(r)}$ & $b_3^{(r)}, b_4^{(r)}$ & $b_5^{(r)}, b_6^{(r)}$ \\
$a_7^{(r)} + b_7^{(r)}$ & $a_8^{(r)} + b_8^{(r)}$ & $a_9^{(r)} + b_9^{(r)}$ \\
\hline
\end{tabular}
\end{center}
\caption{The download scheme for each round $r$, where $r=1$ and $r =2$.}\label{tb:ex1}
\vspace*{-1.0cm}
\end{table}

\noindent {\it Round 1:}
\begin{equation} \label{eqn:ex1_round1_a}
	a_{[1:9]}^{(1)} = \left( {W}_{1}^{[1:7]} \bfG_{[1:7]}^{7 \times 9} +  [ S_1^{(1)} S_2^{(1)}] \bfG_{[8:9]}^{2 \times 9} \right) \bfS_1  
\end{equation}
\begin{equation}\label{eqn:ex1_round1_b}
	b_{[1:9]}^{(1)} = \left(  {W}_{2}^{[8:13]} \bfG_{[1:6]}^{6 \times 9} +  [S_3^{(1)}  S_4^{(1)}  S_5^{(1)}] \bfG_{[7:9]}^{3 \times 9} \right) \bfS_2[:, (1:6)] \bfG_1^{6 \times 9} 
\end{equation}

\noindent {\it Round 2:}
\begin{equation}\label{eqn:ex1_round2_a}
	a_{[1:9]}^{(2)} = \left(  {W}_{1}^{[8:13]} \bfG_{[1:6]}^{6 \times 9} +  [S_3^{(2)}  S_4^{(2)}  S_5^{(2)}] \bfG_{[7:9]}^{3 \times 9} \right) \bfS_3
\end{equation}
\begin{equation}\label{eqn:ex1_round2_b}
	b_{[1:9]}^{(2)} = \left( {W}_{2}^{[1:7]} \bfG_{[1:7]}^{7 \times 9} +  [ S_1^{(2)} S_2^{(2)}] \bfG_{[8:9]}^{2 \times 9} \right) \bfS_4[:, (1:6)] \bfG_1^{6 \times 9} 
\end{equation}

\noindent {\bf Correctness:}
In round 1, the user can solve $b_{[7:9]}^{(1)}$ from $b_{[1:6]}^{(1)}$, because $\bfG_3^{6 \times 9} $ is the generating matrix of a $(9,6)$-MDS code. Therefore, the user can cancel the interference $b_{[7:9]}^{(1)}$ and obtain $a_{[7:9]}^{(1)}$.
From $a_{[1:9]}^{(1)}$, the user can solve ${W}_{1}^{[1:7]}$, because $a_{[1:9]}^{(1)} = [{W}_{1}^{[1:7]} S_1^{(1)} S_2^{(1)}] \bfG \bfS_1$, where $\bfG$ and $\bfS_1$ are invertible matrices.
Similarly in round 2, the user can solve ${W}_{1}^{[8:13]}$. Hence, the user can solve all $13$ symbols of $W_1$.

\noindent {\bf User-privacy:}
Any $T=2$ databases may collude and observe the queries composed of $6$ symbols from $a_{[1:9]}^{(r)}$ and $b_{[1:9]}^{(r)}$ for each round. Let $\calI_a , \calI_b$ denote the indices of the symbols observed by the colluding databases,
\begin{align}
& \quad \left(   a_{\calI_a }^{(1)}, a_{\calI_a }^{(2)}, b_{\calI_b }^{(1)}, b_{\calI_b }^{(2)} \right) \\
& = \Big( \big[{W}_{1}^{[1:7]} S_1^{(1)} S_2^{(1)}\big] \bfG \bfS_1[:, \calI_a] , \big[{W}_{1}^{[8:13]} S_3^{(2)} S_4^{(2)} S_5^{(2)}\big] \bfG \bfS_3[:, \calI_a]   , \\
& \qquad \big[{W}_{2}^{[8:13]} S_3^{(1)} S_4^{(1)} S_5^{(1)} \big] \bfG \bfS_2[:, (1:6)] \bfG_1^{6 \times 9}[:, \calI_b],  \big[ {W}_{2}^{[1:7]} S_1^{(2)} S_2^{(2)} \big] \bfG \bfS_4[:, (1:6)] \bfG_1^{6 \times 9}[:, \calI_b] \Big) \\
& \sim \Big( \big[ {W}_{1}^{[1:7]} S_1^{(1)} S_2^{(1)} \big]  \bfS_1[:, (1:6)] ,  \big[ {W}_{1}^{[8:13]} S_3^{(2)} S_4^{(2)} S_5^{(2)} \big]  \bfS_3[:, (1:6)]   , \\
& \qquad \big[ {W}_{2}^{[8:13]} S_3^{(1)} S_4^{(1)} S_5^{(1)} \big]  \bfS_2[:, (1:6)] ,  \big[{W}_{2}^{[1:7]} S_1^{(2)} S_2^{(2)} \big]  \bfS_4[:, (1:6)] \Big).
\end{align}
The two rounds of download can be randomized by the user.
Therefore, the symbols observed by the two databases are obtained by random mappings from linear combinations of $W_1$ and $W_2$ and the random symbols $S_{[1:5]}^{(1)},S_{[1:5]}^{(2)}$ generated by the databases in the same way, where the randomness of the mapping is privately generated by the user and unavailable to the databases. Hence, user-privacy is guaranteed.

\noindent {\bf System-privacy:}
The eavesdropper can tap in on an arbitrary database. Because the scheme is symmetric across the databases, w.l.o.g., assume DB1 is eavesdropped. 
In round 1, from equation~\eqref{eqn:ex1_round1_a} $a_1^{(1)}, a_2^{(1)}$ are constructed by adding linearly independent combinations of $S_1^{(1)}, S_2^{(1)}$. Similarly from equation~\eqref{eqn:ex1_round1_b}, $b_1^{(1)}, b_2^{(1)}, b_7^{(1)} $ are constructed by adding linearly independent combinations of $S_3^{(1)}, S_4^{(1)}, S_5^{(1)}$. 
Specifically, denote the five answers from DB1 in round 1 by $A_{DB1}^{(1)}$, the linear combinations of the $S_{[1:5]}^{(1)}$ added to the answers are constructed by,
\begin{equation}
	[S_1^{(1)} S_2^{(1)} S_3^{(1)} S_4^{(1)} S_5^{(1)} ] \cdot
	\begin{bmatrix}
	\left[ \bfG_{[8:9]}^{2 \times 9} \bfS_1[:, (1:2)] \right]^{2 \times 2} & \mathbf{0}^{2 \times 2} \qquad \left[ \bfG_{[8:9]}^{2 \times 9} \bfS_1[:, 7] \right]^{2 \times 1} \\
	 \mathbf{0}^{3 \times 2}  & \left[ \bfG_{[7:9]}^{2 \times 9} \bfS_2[:, (1:6)] \bfG_1^{6 \times 9} [:,(1,2,7)] \right]^{3 \times 3} 
	\end{bmatrix}. \label{eqn:ex1_sys_privacy}
\end{equation}
It can be checked that the $5 \times 5$ matrix in~\eqref{eqn:ex1_sys_privacy} is invertible. Therefore, $H(A_{DB1}^{(1)}) = H(A_{DB1}^{(1)} | W_1, W_2) = 5 \log{q} $.
Hence, $I (A_{DB1}^{(1)} ; W_1, W_2) =0$.
The construction of symbols for round 2 are in a similar way, by adding linearly independent combinations of $S_{[1:5]}^{(2)}$. Because the $10$ symbols $S_{[1:5]}^{(1)},S_{[1:5]}^{(2)}$ are independently and uniformly chosen from $\Fq$, we have $I (A_{DB1}^{(1)} , A_{DB1}^{(2)} ; W_1, W_2) =0$ and hence the eavesdropper obtains no information regarding the database $W_1, W_2$.

\subsection{Example: $N = 4$ databases, $K = 2$ files, $T = 2$ colluding databases, $E = 1$ eavesdropped database}
Suppose each file consists of $L = 13$ symbols and is represented as a length-$13$ row vector over a sufficiently large field $\Fq$, denoted by $W_1 = W_1^{[1:13]}$ and $W_2 = W_2^{[1:13]}$. The user downloads two rounds. For each round, the user downloads $12$ symbols. The databases generate $6$ uniformly random symbols $S_{[1:3]}^{(1)}, S_{[1:3]}^{(2)}$. 
The scheme achieves the rate $R = 13/24$.


The user privately generates matrices  $\bfS_1, \bfS_2  ,\bfS_3, \bfS_4 \in \Fq^{8 \times 8}$ uniformly and independently from all $8 \times 8$ invertible matrices over $\Fq$.

Let $\{ \lambda_1, \dots, \lambda_8 \}$ be $8$ distinct nonzero elements from $\Fq$. Let $\bfG$ be a $8 \times 8 $ matrix defined as follows,
\begin{equation}
{
\bfG= 
\begin{bmatrix}
1 & 1 & \dots & 1\\
\lambda_1 & \lambda_2 & \dots & \lambda_8       \\
\vdots  & \vdots   & \ddots & \vdots   \\
\lambda_1^{7} & \lambda_2^{7} &  \dots & \lambda_8^{7}   
\end{bmatrix}
}
, 
\end{equation}
it is direct that $\bfG$ is an invertible matrix. Let $\bfG_{[1:7]}^{7 \times 8}$ and $\bfG_{[8]}^{1 \times 8}$ be matrices composed of the first $7$ rows and the $8$th row respectively, such that $\bfG = [\bfG_{[1:7]}^{7 \times 8} \;  \bfG_{[8]}^{1 \times 8} ]^{\textrm{T}}$. Similarly, let $\bfG_{[1:6]}^{6 \times 8}$ and $\bfG_{[7:8]}^{2 \times 8}$ be matrices composed of the first $6$ rows and the last $2$ rows respectively, such that $\bfG = [\bfG_{[1:6]}^{6 \times 8} \;  \bfG_{[7:8]}^{2 \times 8} ]^{\textrm{T}}$. The matrices $\bfG_{[1:7]}^{7 \times 8}$, $\bfG_{[8]}^{1 \times 8}$, $\bfG_{[1:6]}^{6 \times 8}$ and $\bfG_{[7:8]}^{2 \times 8}$ are generating matrices of MDS codes with corresponding dimensions.

Let $\bfG_1^{4 \times 8}$ be the generating matrix of a $(8,4)$-MDS code.

\begin{center}
\begin{tabular}{|c|c|c|c|}
\hline
DB1 & DB2 & DB3 & DB4 \\
\hline
$a_1^{(r)}$ & $a_2^{(r)}$ & $a_3^{(r)}$ & $a_4^{(r)}$ \\
$b_1^{(r)}$ & $b_2^{(r)}$ & $b_3^{(r)}$ & $b_4^{(r)}$ \\
$a_5^{(r)} + b_5^{(r)}$ & $a_6^{(r)} + b_6^{(r)}$ & $a_7^{(r)} + b_7^{(r)}$ & $a_8^{(r)} + b_8^{(r)}$ \\
\hline
\end{tabular}
\end{center}

\noindent {\it Round 1:}
\begin{equation}
	a_{[1:8]}^{(1)} = \left( {W}_{1}^{[1:7]} \bfG_{[1:7]}^{7 \times 8} +  S_1^{(1)} \bfG_{[8]}^{1 \times 8} \right) \bfS_1
\end{equation}
\begin{equation}
	b_{[1:8]}^{(1)} = \left(  {W}_{2}^{[8:13]} \bfG_{[1:6]}^{6 \times 8} +  [S_2^{(1)}  S_3^{(1)}  ] \bfG_{[7:8]}^{2 \times 8} \right) \bfS_2[:, (1:4)] \bfG_1^{4 \times 8} 
\end{equation}

\noindent {\it Round 2:}
\begin{equation}
	a_{[1:8]}^{(2)} = \left(  {W}_{1}^{[8:13]} \bfG_{[1:6]}^{6 \times 8} +  [S_2^{(2)}  S_3^{(2)}  ] \bfG_{[7:8]}^{2 \times 8} \right) \bfS_3
\end{equation}
\begin{equation}
	b_{[1:8]}^{(2)} = \left( {W}_{2}^{[1:7]} \bfG_{[1:7]}^{7 \times 8} +  S_1^{(2)}  \bfG_{[8]}^{1 \times 8} \right) \bfS_4[:, (1:4)] \bfG_1^{4 \times 8} 
\end{equation}

\subsection{Example: $N = 4$ databases, $K = 2$ files, $T = 3$ colluding databases, $E = 1$ eavesdropped database}
Suppose each file consists of $L = 25$ symbols and is represented as a length-$25$ row vector over a sufficiently large field $\Fq$, denoted by $W_1 = W_1^{[1:25]}$ and $W_2 = W_2^{[1:25]}$. The user downloads two rounds. For each round, the user downloads $28$ symbols. The databases generate $14$ uniformly random symbols $S_{[1:7]}^{(1)}, S_{[1:7]}^{(2)}$. 
The scheme achieves the rate $R = 25/56$.


The user privately generates matrices  $\bfS_1, \bfS_2  ,\bfS_3, \bfS_4 \in \Fq^{16 \times 16}$ uniformly and independently from all $16 \times 16$ invertible matrices over $\Fq$.

Let $\{ \lambda_1, \dots, \lambda_{16} \}$ be $16$ distinct nonzero elements from $\Fq$. Let $\bfG$ be a $16 \times 16$ matrix defined as follows,
\begin{equation}
{
\bfG_2= 
\begin{bmatrix}
1 & 1 & \dots & 1\\
\lambda_1 & \lambda_2 & \dots & \lambda_{16}       \\
\vdots  & \vdots   & \ddots & \vdots   \\
\lambda_1^{15} & \lambda_2^{15} &  \dots & \lambda_{16}^{15}   
\end{bmatrix}
}
, 
\end{equation}
it is direct that $\bfG$ is an invertible matrix. Let $\bfG_{[1:13]}^{13 \times 16}$ and $\bfG_{[14:16]}^{3 \times 16}$ be matrices composed of the first $13$ rows and the last $3$ rows respectively, such that $\bfG = [\bfG_{[1:13]}^{13 \times 16} \;  \bfG_{[14:16]}^{3 \times 16} ]^{\textrm{T}}$. Similarly, let $\bfG_{[1:12]}^{12 \times 16}$ and $\bfG_{[13:16]}^{4 \times 16}$ be matrices composed of the first $12$ rows and the last $4$ rows respectively, such that $\bfG = [\bfG_{[1:12]}^{12 \times 16} \;  \bfG_{[13:16]}^{4 \times 16} ]^{\textrm{T}}$. The matrices $\bfG_{[1:13]}^{13 \times 16}$, $\bfG_{[14:16]}^{3 \times 16}$, $\bfG_{[1:12]}^{12 \times 16}$ and $\bfG_{[13:16]}^{4 \times 16}$ are generating matrices of MDS codes with corresponding dimensions.

Let $\bfG_1^{12 \times 16}$ be the generating matrix of a $(16,12)$-MDS code.

\begin{center}
\begin{tabular}{|c|c|c|c|}
\hline
DB1 & DB2 & DB3 & DB4 \\
\hline
$a_1^{(r)},a_2^{(r)},a_3^{(r)}$ & $a_4^{(r)},a_5^{(r)},a_6^{(r)}$ & $a_7^{(r)},a_8^{(r)},a_9^{(r)}$ & $a_{10}^{(r)},a_{11}^{(r)},a_{12}^{(r)}$ \\
$b_1^{(r)},b_2^{(r)},b_3^{(r)}$ & $b_4^{(r)},b_5^{(r)},b_6^{(r)}$ & $b_7^{(r)},b_8^{(r)},b_9^{(r)}$ & $b_{10}^{(r)},b_{11}^{(r)},b_{12}^{(r)}$ \\
$a_{13}^{(r)} + b_{13}^{(r)}$ & $a_{14}^{(r)} + b_{14}^{(r)}$ & $a_{15}^{(r)} + b_{15}^{(r)}$ & $a_{16}^{(r)} + b_{16}^{(r)}$ \\
\hline
\end{tabular}
\end{center}

\noindent {\it Round 1:}
\begin{equation}
	a_{[1:16]}^{(1)} = \left( {W}_{1}^{[1:13]} \bfG_{[1:13]}^{13 \times 16} +  [ S_1^{(1)} S_2^{(1)} S_3^{(1)}]  \bfG_{[14:16]}^{3 \times 16} \right) \bfS_1
\end{equation}
\begin{equation}
	b_{[1:16]}^{(1)} = \left(  {W}_{2}^{[14:25]} \bfG_{[1:12]}^{12 \times 16} +  [S_4^{(1)}  S_5^{(1)} S_6^{(1)}  S_7^{(1)}  ] \bfG_{[13:16]}^{4 \times 16} \right) \bfS_2[:, (1:12)] \bfG_1^{12 \times 16} 
\end{equation}

\noindent {\it Round 2:}
\begin{equation}
	a_{[1:16]}^{(2)} = \left(  {W}_{1}^{14:25]} \bfG_{[1:12]}^{12 \times 16} +  [S_4^{(2)}  S_5^{(2)} S_6^{(2)}  S_7^{(2)}  ] \bfG_{[13:16]}^{4 \times 16} \right) \bfS_3
\end{equation}
\begin{equation}
	b_{[1:16]}^{(2)} = \left( {W}_{2}^{[1:13]} \bfG_{[1:13]}^{13 \times 16} +  [  S_1^{(2)} S_2^{(2)}  S_3^{(2)}  ]  \bfG_{[14:16]}^{3 \times 16} \right) \bfS_4[:, (1:12)] \bfG_1^{12 \times 16} 
\end{equation}

\subsection{Example: $N = 4$ databases, $K = 2$ files, $T = 3$ colluding databases, $E = 2$ eavesdropped databases}
Suppose each file contains $L = 18$ symbols and is represented as a length-$18$ row vector over a sufficiently large field $\Fq$, denoted by $W_1 = W_1^{[1:18]}$ and $W_2 = W_2^{[1:18]}$. The user downloads two rounds. For each round, the user downloads $28$ symbols. The databases generate $28$ uniformly random symbols $S_{[1:14]}^{(1)}, S_{[1:14]}^{(2)}$. 
The scheme achieves the rate $R = 18/56 = 9/28$.


The user privately generates matrices  $\bfS_1, \bfS_2  ,\bfS_3, \bfS_4 \in \Fq^{16 \times 16}$ uniformly and independently from all $16 \times 16$ invertible matrices over $\Fq$.

Let $\{ \lambda_1, \dots, \lambda_{16} \}$ be $16$ distinct nonzero elements from $\Fq$. Let $\bfG$ be a $16 \times 16$ matrix defined as follows,
\begin{equation}
{
\bfG= 
\begin{bmatrix}
1 & 1 & \dots & 1\\
\lambda_1 & \lambda_2 & \dots & \lambda_{16}       \\
\vdots  & \vdots   & \ddots & \vdots   \\
\lambda_1^{15} & \lambda_2^{15} &  \dots & \lambda_{16}^{15}   
\end{bmatrix}
}
, 
\end{equation}
it is direct that $\bfG$ is an invertible matrix. Let $\bfG_{[1:10]}^{10 \times 16}$ and $\bfG_{[11:16]}^{6 \times 16}$ be matrices composed of the first $10$ rows and the last $6$ rows respectively, such that $\bfG = [\bfG_{[1:10]}^{10 \times 16} \;  \bfG_{[11:16]}^{6 \times 16} ]^{\textrm{T}}$. Similarly, let $\bfG_{[1:8]}^{8 \times 16}$ and $\bfG_{[9:16]}^{8 \times 16}$ be matrices composed of the first $8$ rows and the last $8$ rows respectively, such that $\bfG = [\bfG_{[1:8]}^{8 \times 16} \;  \bfG_{[9:16]}^{8 \times 16} ]^{\textrm{T}}$. The matrices $\bfG_{[1:10]}^{10 \times 16}$, $\bfG_{[11:16]}^{6 \times 16}$, $\bfG_{[1:8]}^{8 \times 16}$ and $\bfG_{[9:16]}^{8 \times 16}$ are generating matrices of MDS codes with corresponding dimensions.

Let $\bfG_1^{12 \times 16}$ be the generating matrix of a $(16,12)$-MDS code.

\begin{center}
\begin{tabular}{|c|c|c|c|}
\hline
DB1 & DB2 & DB3 & DB4 \\
\hline
$a_1^{(r)},a_2^{(r)},a_3^{(r)}$ & $a_4^{(r)},a_5^{(r)},a_6^{(r)}$ & $a_7^{(r)},a_8^{(r)},a_9^{(r)}$ & $a_{10}^{(r)},a_{11}^{(r)},a_{12}^{(r)}$ \\
$b_1^{(r)},b_2^{(r)},b_3^{(r)}$ & $b_4^{(r)},b_5^{(r)},b_6^{(r)}$ & $b_7^{(r)},b_8^{(r)},b_9^{(r)}$ & $b_{10}^{(r)},b_{11}^{(r)},b_{12}^{(r)}$ \\
$a_{13}^{(r)} + b_{13}^{(r)}$ & $a_{14}^{(r)} + b_{14}^{(r)}$ & $a_{15}^{(r)} + b_{15}^{(r)}$ & $a_{16}^{(r)} + b_{16}^{(r)}$ \\
\hline
\end{tabular}
\end{center}

\noindent {\it Round 1:}
\begin{equation}
	a_{[1:16]}^{(1)} = \left( {W}_{1}^{[1:10]} \bfG_{[1:10]}^{10 \times 16} +  S_{[1:6]}^{(1)}   \bfG_{[11:16]}^{6 \times 16} \right) \bfS_1
\end{equation}
\begin{equation}
	b_{[1:16]}^{(1)} = \left(  {W}_{2}^{[11:18]} \bfG_{[1:8]}^{8 \times 16} +  S_{[7:14]}^{(1)}   \bfG_{[9:16]}^{8 \times 16} \right) \bfS_2[:, (1:12)] \bfG_1^{12 \times 16} 
\end{equation}

\noindent {\it Round 2:}
\begin{equation}
	a_{[1:16]}^{(2)} = \left(  {W}_{1}^{[11:18]} \bfG_{[1:8]}^{8 \times 16} +  S_{[7:14]}^{(2)}   \bfG_{[9:16]}^{8 \times 16} \right) \bfS_3
\end{equation}
\begin{equation}
	b_{[1:16]}^{(2)} = \left( {W}_{2}^{[1:10]} \bfG_{[1:10]}^{10 \times 16} +    S_{[1:6]}^{(2)}  \bfG_{[11:16]}^{6 \times 16} \right) \bfS_4[:, (1:12)] \bfG_1^{12 \times 16} 
\end{equation}

\subsection{Example: $N = 3$ databases, $K = 3$ files, $T = 2$ colluding databases, $E = 1$ eavesdropped database} \label{sec:ex5}
Suppose each file contains $L = 62$ symbols. Let the symbols of each file be randomly permuted (the randomness is generated privately by the user) and be represented as a length-$62$ row vector over a sufficiently large field $\Fq$, denoted by $W_1 = W_1^{[1:62]}$, $W_2 = W_2^{[1:62]}$ and $W_3 = W_3^{[1:62]}$. The user downloads three rounds. For each round, the user downloads $57$ symbols. The databases generate $57$ uniformly random symbols, $19$ for each round and denoted as $S_{[1:19]}^{(1)}, S_{[1:19]}^{(2)},S_{[1:19]}^{(3)}$, for protecting the database from the eavesdropper. 
The scheme achieves the rate $R = 62/171$.


Let $\{ \lambda_1, \dots, \lambda_{27} \}$ be $27$ distinct nonzero elements from $\Fq$. Let $\bfG$ be a $27 \times 27$ matrix defined as follows,
\begin{equation}
{
\bfG= 
\begin{bmatrix}
1 & 1 & \dots & 1\\
\lambda_1 & \lambda_2 & \dots & \lambda_{27}       \\
\vdots  & \vdots   & \ddots & \vdots   \\
\lambda_1^{26} & \lambda_2^{26} &  \dots & \lambda_{27}^{26}   
\end{bmatrix}
}
, 
\end{equation}
it is direct that $\bfG$ is an invertible matrix. Let $\bfG_{[1:18]}^{18 \times 27}$ and $\bfG_{[19:27]}^{9 \times 27}$ be matrices composed of the first $18$ rows and the last $9$ rows respectively, such that $\bfG = [\bfG_{[1:18]}^{18 \times 27} \;  \bfG_{[19:27]}^{9 \times 27} ]^{\textrm{T}}$. Similarly, let $\bfG_{[1:21]}^{21 \times 27}$ and $\bfG_{[22:27]}^{6 \times 27}$ be matrices composed of the first $21$ rows and the last $6$ rows respectively, and $\bfG_{[1:23]}^{23 \times 27}$ and $\bfG_{[24:27]}^{4 \times 27}$ be matrices composed of the first $23$ rows and the last $4$ rows respectively. The matrices $\bfG_{[1:18]}^{18 \times 27}$, $\bfG_{[19:27]}^{9 \times 27}$, $\bfG_{[1:21]}^{21 \times 27}$ ,$\bfG_{[22:27]}^{6 \times 27}$, $\bfG_{[1:23]}^{23 \times 27}$ and $\bfG_{[24:27]}^{4 \times 27}$ are generating matrices of MDS codes with corresponding dimensions.

The user privately generates $9$ matrices  $\bfS_{[1:3]}^{(1)},\bfS_{[1:3]}^{(2)},\bfS_{[1:3]}^{(3)} \in \Fq^{27 \times 27}$ uniformly and independently from all $27 \times 27$ invertible matrices over $\Fq$.

Let $\bfG_1^{12 \times 18}$ and  $\bfG_2^{6 \times 9}$ be the generating matrices of a $(18,12)$-MDS code and a $(9,6)$-MDS code respectively.

\begin{center}
\begin{tabular}{|c|c|c|}
\hline
DB1 & DB2 & DB3  \\
\hline
$a_1^{(r)},a_2^{(r)},a_3^{(r)},a_4^{(r)}$ & $a_5^{(r)},a_6^{(r)},a_7^{(r)},a_8^{(r)}$ & $a_9^{(r)},a_{10}^{(r)},a_{11}^{(r)},a_{12}^{(r)}$  \\
$b_1^{(r)},b_2^{(r)},b_3^{(r)},b_4^{(r)}$ & $b_5^{(r)},b_6^{(r)},b_7^{(r)},b_8^{(r)}$ & $b_9^{(r)},b_{10}^{(r)},b_{11}^{(r)},b_{12}^{(r)}$  \\
$c_1^{(r)},c_2^{(r)},c_3^{(r)},c_4^{(r)}$ & $c_5^{(r)},c_6^{(r)},c_7^{(r)},c_8^{(r)}$ & $c_9^{(r)},c_{10}^{(r)},c_{11}^{(r)},c_{12}^{(r)}$  \\
$a_{13}^{(r)} + b_{13}^{(r)}$ & $a_{15}^{(r)} + b_{15}^{(r)}$ & $a_{21}^{(r)} + b_{17}^{(r)}$ \\
$a_{14}^{(r)} + b_{14}^{(r)}$ & $a_{16}^{(r)} + b_{16}^{(r)}$ & $a_{22}^{(r)} + b_{18}^{(r)}$ \\
$a_{17}^{(r)} + c_{13}^{(r)}$ & $a_{19}^{(r)} + c_{15}^{(r)}$ & $a_{23}^{(r)} + c_{17}^{(r)}$ \\
$a_{18}^{(r)} + c_{14}^{(r)}$ & $a_{20}^{(r)} + c_{16}^{(r)}$ & $a_{24}^{(r)} + c_{18}^{(r)}$ \\
$b_{19}^{(r)} + c_{19}^{(r)}$ & $b_{21}^{(r)} + c_{21}^{(r)}$ & $b_{23}^{(r)} + c_{23}^{(r)}$ \\
$b_{20}^{(r)} + c_{20}^{(r)}$ & $b_{22}^{(r)} + c_{22}^{(r)}$ & $b_{24}^{(r)} + c_{24}^{(r)}$ \\
$a_{25}^{(r)} +b_{25}^{(r)} + c_{25}^{(r)}$ & $a_{26}^{(r)} +b_{26}^{(r)} + c_{26}^{(r)}$ & $a_{27}^{(r)} +b_{27}^{(r)} + c_{27}^{(r)}$ \\
\hline
\end{tabular}
\end{center}

\noindent {\it Round 1:}
\begin{equation}
	a_{[1:27]}^{(1)} = \left( {W}_{1}^{[1:18]}  \bfG_{[1:18]}^{18 \times 27} +  S_{[1:9]}^{(1)}  \bfG_{[19:27]}^{9 \times 27} \right) \bfS_1^{(1)}
\end{equation}
\begin{equation}
	b_{[1:18]}^{(1)} = \left(  {W}_{2}^{[19:39]} \bfG_{[1:21]}^{21 \times 27} +  S_{[10:15]}^{(1)}   \bfG_{[22:27]}^{6 \times 27} \right) \bfS_2^{(1)}[:, (1:12)] \bfG_1^{12 \times 18} 
\end{equation}
\begin{equation}
	b_{[19:27]}^{(1)} = \left(  {W}_{2}^{[19:39]} \bfG_{[1:21]}^{21 \times 27} +  S_{[10:15]}^{(1)}   \bfG_{[22:27]}^{6 \times 27} \right) \bfS_2^{(1)}[:, (13:18)] \bfG_2^{6 \times 9} 
\end{equation}
\begin{equation}
	c_{[1:18]}^{(1)} = \left(  {W}_{3}^{[40:62]} \bfG_{[1:23]}^{23 \times 27} +  S_{[16:19]}^{(1)}   \bfG_{[24:27]}^{4 \times 27} \right) \bfS_3^{(1)}[:, (1:12)] \bfG_1^{12 \times 18} 
\end{equation}
\begin{equation}
	c_{[19:27]}^{(1)} = \left(  {W}_{3}^{[40:62]} \bfG_{[1:23]}^{23 \times 27} +  S_{[16:19]}^{(1)}   \bfG_{[24:27]}^{4 \times 27} \right) \bfS_3^{(1)}[:, (13:18)]  \bfG_2^{6 \times 9} 
\end{equation}

\noindent {\it Round 2:}
\begin{equation}
	a_{[1:27]}^{(2)} = \left( {W}_{1}^{[19:39]}  \bfG_{[1:21]}^{21 \times 27} +  S_{[10:15]}^{(2)}  \bfG_{[22:27]}^{6 \times 27} \right) \bfS_1^{(2)}
\end{equation}
\begin{equation}
	b_{[1:18]}^{(2)} = \left(  {W}_{2}^{[40:62]} \bfG_{[1:23]}^{23 \times 27} +  S_{[16:19]}^{(2)}   \bfG_{[24:27]}^{4 \times 27} \right) \bfS_2^{(2)}[:, (1:12)] \bfG_1^{12 \times 18} 
\end{equation}
\begin{equation}
	b_{[19:27]}^{(1)} = \left(  {W}_{2}^{[40:62]} \bfG_{[1:23]}^{23 \times 27} +  S_{[16:19]}^{(2)}   \bfG_{[24:27]}^{4 \times 27} \right) \bfS_2^{(2)}[:, (13:18)]  \bfG_2^{6 \times 9} 
\end{equation}
\begin{equation}
	c_{[1:18]}^{(1)} = \left(  {W}_{3}^{[1:18]} \bfG_{[1:18]}^{18 \times 27} +  S_{[1:9]}^{(2)}   \bfG_{[19:27]}^{9 \times 27} \right) \bfS_3^{(2)}[:, (1:12)] \bfG_1^{12 \times 18} 
\end{equation}
\begin{equation}
	c_{[19:27]}^{(1)} = \left(  {W}_{3}^{[1:18]} \bfG_{[1:18]}^{18 \times 27} +  S_{[1:9]}^{(2)}   \bfG_{[19:27]}^{9 \times 27} \right) \bfS_3^{(2)}[:, (13:18)]  \bfG_2^{6 \times 9} 
\end{equation}

\noindent {\it Round 3:}
\begin{equation}
	a_{[1:27]}^{(3)} = \left( {W}_{1}^{[40:62]}  \bfG_{[1:23]}^{23 \times 27} +  S_{[16:19]}^{(3)}  \bfG_{[29:27]}^{4 \times 27} \right) \bfS_1^{(3)}
\end{equation}
\begin{equation}
	b_{[1:18]}^{(3)} = \left(  {W}_{1}^{[1:18]} \bfG_{[1:18]}^{18 \times 27} +  S_{[1:9]}^{(3)}   \bfG_{[19:27]}^{9 \times 27} \right) \bfS_2^{(3)}[:, (1:12)] \bfG_1^{12 \times 18}
\end{equation}
\begin{equation}
	b_{[19:27]}^{(3)} = \left(  {W}_{1}^{[1:18]} \bfG_{[1:18]}^{18 \times 27} +  S_{[1:9]}^{(3)}   \bfG_{[19:27]}^{9 \times 27} \right) \bfS_2^{(3)}[:, (13:18)] \bfG_2^{6 \times 9}
\end{equation}
\begin{equation}
	c_{[1:18]}^{(3)} = \left(  {W}_{3}^{[19:39]} \bfG_{[1:21]}^{21 \times 27} +  S_{[10:15]}^{(3)}   \bfG_{[22:27]}^{6 \times 27} \right) \bfS_3^{(3)}[:, (1:12)] \bfG_1^{12 \times 18}
\end{equation}
\begin{equation}
	c_{[19:27]}^{(3)} = \left(  {W}_{3}^{[19:39]} \bfG_{[1:21]}^{21 \times 27} +  S_{[10:15]}^{(3)}   \bfG_{[22:27]}^{6 \times 27} \right) \bfS_3^{(3)}[:, (13:18)]  \bfG_2^{6 \times 9} 
\end{equation}

\noindent {\bf Correctness:}
For each round, the user can recover $b_{[13:18]}^{(r)}$ and $c_{[13:18]}^{(r)}$ from $b_{[1:12]}^{(r)}$ and $c_{[1:12]}^{(r)}$. Therefore, the user can cancel the interference and solve $a_{[13:24]}^{(r)}$.
Similarly, the user can recover and cancel $b_{[25:27]}^{(r)}+c_{[25:27]}^{(r)}$ and obtain $a_{[25:27]}^{(r)}$, because $b_{[19:27]}^{(r)}$ and $c_{[19:27]}^{(r)}$ are generated from the same $(9,6)$-MDS code.
Hence, the user can solve $a_{[1:27]}^{(r)}$ for all three rounds.
For round 1, we have$a_{[1:27]}^{(1)} = \big[ {W}_{1}^{[1:18]}    S_{[1:9]}^{(1)}  \big] \bfG \bfS_1^{(1)}$. Because $\bfG$ and $\bfS_1^{(1)}$ are invertible matrices, the user can solve $18$ symbols ${W}_{1}^{[1:18]}$. Similarly, the user can solve ${W}_{1}^{[19:39]}$ and ${W}_{1}^{[40:62]}$ for both round 2 and round 3.
Hence, the user obtains all $62$ symbols of $W_1$.

\noindent {\bf User-privacy:}
Any $T=2$ databases may collude and observe the queries composed of $18$ symbols from $a_{[1:27]}^{(r)}$, $12$ symbols from both $b_{[1:18]}^{(r)}$ and $c_{[1:18]}^{(r)}$, and $6$ symbols from both $b_{[19:27]}^{(r)}$ and $c_{[19:27]}^{(r)}$ for each round. Let $\calI_a , \calI_{b,12}, \calI_{b,6}, \calI_{c,12}, \calI_{c,6}$ denote the indices of the symbols observed by the colluding databases,
\begin{align}
& \quad
\begin{pmatrix}
a_{\calI_a}^{(1)}, & a_{\calI_a}^{(2)}, & a_{\calI_a}^{(3)} \\
(b_{\calI_{b,12}}^{(1)}, b_{\calI_{b,6}}^{(1)}),   & (b_{\calI_{b,12}}^{(2)}, b_{\calI_{b,6}}^{(2)}),   & (b_{\calI_{b,12}}^{(3)}, b_{\calI_{b,6}}^{(3)}) \\
(c_{\calI_{c,12}}^{(1)}, c_{\calI_{c,6}}^{(1)}) , &  (c_{\calI_{c,12}}^{(2)}, c_{\calI_{c,6}}^{(2)}) ,   & (c_{\calI_{c,12}}^{(3)}, c_{\calI_{c,6}}^{(3)}) 
\end{pmatrix}
\\
& = 
\begin{pmatrix}
\big[ {W}_{1}^{[1:18]}    S_{[1:9]}^{(1)}  \big] \bfG \bfS_1^{(1)} [:, \calI_a],  \big[ {W}_{1}^{[19:39]}    S_{[10:15]}^{(2)}  \big] \bfG \bfS_1^{(2)} [:, \calI_a],  \big[ {W}_{1}^{[40:62]}    S_{[16:19]}^{(3)}  \big] \bfG \bfS_1^{(3)} [:, \calI_a] \\
\begin{bmatrix}
 (\big[ {W}_{2}^{[19:39]}    S_{[10:15]}^{(1)}  \big] \bfG \bfS_2^{(1)} [:, (1:12)] \bfG_1^{12 \times 18} [:, \calI_{b,12}], \big[ {W}_{2}^{[19:39]}    S_{[10:15]}^{(1)}  \big] \bfG \bfS_2^{(1)} [:, (13:18)] \bfG_2^{6 \times 9} [:, \calI_{b,6}] ) \\
    (\big[ {W}_{2}^{[40:62]}    S_{[16:19]}^{(2)}  \big] \bfG \bfS_2^{(2)} [:, (1:12)] \bfG_1^{12 \times 18} [:, \calI_{b,12}], \big[ {W}_{2}^{[40:62]}    S_{[16:19]}^{(2)}  \big] \bfG \bfS_2^{(2)} [:, (13:18)] \bfG_2^{6 \times 9} [:, \calI_{b,6}] )   \\
     (\big[ {W}_{2}^{[1:18]}    S_{[1:9]}^{(3)}  \big] \bfG \bfS_2^{(3)} [:, (1:12)] \bfG_1^{12 \times 18} [:, \calI_{b,12}], \big[ {W}_{2}^{[1:18]}    S_{[1:9]}^{(3)}  \big] \bfG \bfS_2^{(3)} [:, (13:18)] \bfG_2^{6 \times 9} [:, \calI_{b,6}] ) 
\end{bmatrix}^{\textrm{T}} \\
\begin{bmatrix}
 (\big[ {W}_{3}^{[40:62]}    S_{[16:19]}^{(1)}  \big] \bfG \bfS_3^{(1)} [:, (1:12)] \bfG_1^{12 \times 18} [:, \calI_{c,12}], \big[ {W}_{3}^{[40:62]}    S_{[16:19]}^{(1)}  \big] \bfG \bfS_3^{(1)} [:, (13:18)] \bfG_2^{6 \times 9} [:, \calI_{c,6}] )   \\
     (\big[ {W}_{3}^{[1:18]}    S_{[1:9]}^{(2)}  \big] \bfG \bfS_3^{(2)} [:, (1:12)] \bfG_1^{12 \times 18} [:, \calI_{c,12}], \big[ {W}_{3}^{[1:18]}    S_{[1:9]}^{(2)}  \big] \bfG \bfS_3^{(2)} [:, (13:18)] \bfG_2^{6 \times 9} [:, \calI_{c,6}] ) \\
     (\big[ {W}_{3}^{[19:39]}    S_{[10:15]}^{(3)}  \big] \bfG \bfS_3^{(3)} [:, (1:12)] \bfG_1^{12 \times 18} [:, \calI_{c,12}], \big[ {W}_{3}^{[19:39]}    S_{[10:15]}^{(3)}  \big] \bfG \bfS_3^{(3)} [:, (13:18)] \bfG_2^{6 \times 9} [:, \calI_{c,6}] ) \\
\end{bmatrix}^{\textrm{T}}
\end{pmatrix} \\
& \sim
\begin{pmatrix}
\big[ {W}_{1}^{[1:18]}    S_{[1:9]}^{(1)}  \big]  \bfG \bfS_1^{(1)} [:, (1:18)],  \big[ {W}_{1}^{[19:39]}    S_{[10:15]}^{(2)}  \big]  \bfG \bfS_1^{(2)} [:, (1:18)],  \big[ {W}_{1}^{[40:62]}    S_{[16:19]}^{(3)}  \big] \bfG  \bfS_1^{(3)} [:, (1:18)] \\
\begin{bmatrix}
 (\big[ {W}_{2}^{[19:39]}    S_{[10:15]}^{(1)}  \big]  \bfG \bfS_2^{(1)} [:, (1:12)] , \big[ {W}_{2}^{[19:39]}    S_{[10:15]}^{(1)}  \big] \bfG \bfS_2^{(1)} [:, (13:18)] ) \\
    (\big[ {W}_{2}^{[40:62]}    S_{[16:19]}^{(2)}  \big]  \bfG \bfS_2^{(2)} [:, (1:12)] , \big[ {W}_{2}^{[40:62]}    S_{[16:19]}^{(2)}  \big] \bfG  \bfS_2^{(2)} [:, (13:18)]  )   \\
     (\big[ {W}_{2}^{[1:18]}    S_{[1:9]}^{(3)}  \big] \bfG \bfS_2^{(3)} [:, (1:12)] , \big[ {W}_{2}^{[1:18]}    S_{[1:9]}^{(3)}  \big]  \bfG \bfS_2^{(3)} [:, (13:18)]  ) 
\end{bmatrix}^{\textrm{T}} \\
\begin{bmatrix}
 (\big[ {W}_{3}^{[40:62]}    S_{[16:19]}^{(1)}  \big] \bfG \bfS_3^{(1)} [:, (1:12)] , \big[ {W}_{3}^{[40:62]}    S_{[16:19]}^{(1)}  \big]  \bfG \bfS_3^{(1)} [:, (13:18)] )   \\
     (\big[ {W}_{3}^{[1:18]}    S_{[1:9]}^{(2)}  \big]  \bfG \bfS_3^{(2)} [:, (1:12)] , \big[ {W}_{3}^{[1:18]}    S_{[1:9]}^{(2)}  \big]  \bfG \bfS_3^{(2)} [:, (13:18)]  ) \\
     (\big[ {W}_{3}^{[19:39]}    S_{[10:15]}^{(3)}  \big]  \bfG \bfS_3^{(3)} [:, (1:12)] , \big[ {W}_{3}^{[19:39]}    S_{[10:15]}^{(3)}  \big] \bfG \bfS_3^{(3)} [:, (13:18)]  ) \\
\end{bmatrix}^{\textrm{T}}
\end{pmatrix} \\
& = 
\begin{pmatrix}
\big[ {W}_{1}^{[1:18]}    S_{[1:9]}^{(1)}  \big]  \bfG \bfS_1^{(1)} [:, (1:18)],  \big[ {W}_{1}^{[19:39]}    S_{[10:15]}^{(2)}  \big]  \bfG \bfS_1^{(2)} [:, (1:18)],  \big[ {W}_{1}^{[40:62]}    S_{[16:19]}^{(3)}  \big] \bfG \bfS_1^{(3)} [:, (1:18)] \\
\big[ {W}_{2}^{[19:39]}    S_{[10:15]}^{(1)}  \big] \bfG  \bfS_2^{(1)} [:, (1:18)] , \big[ {W}_{2}^{[40:62]}    S_{[16:19]}^{(2)}  \big] \bfG  \bfS_2^{(2)} [:, (1:18)] , \big[ {W}_{2}^{[1:18]}    S_{[1:9]}^{(3)}  \big] \bfG  \bfS_2^{(3)} [:, (1:18)] \\
\big[ {W}_{3}^{[40:62]}    S_{[16:19]}^{(1)}  \big] \bfG \bfS_3^{(1)} [:, (1:18)] , \big[ {W}_{3}^{[1:18]}    S_{[1:9]}^{(2)}  \big]  \bfG \bfS_3^{(2)} [:, (1:18)] , \big[ {W}_{3}^{[19:39]}    S_{[10:15]}^{(3)}  \big] \bfG  \bfS_3^{(3)} [:, (1:18)] 
\end{pmatrix} 
\end{align}
The user can randomize the three rounds of downloading. Therefore, the symbols requested at the two colluding databases are mapped from the symbols of each file and the $S_i^{(r)}$'s in the same way. Hence, user-privacy is guaranteed.

\noindent {\bf System-privacy:}
Similar as in the example in Section~\ref{sec:ex1}, the answers from any database is composed by adding linearly independent combinations of $S_{[1:19]}^{(r)}$ for each round. Therefore, the eavesdropper obtains no information regarding the database $W_1, W_2, W_3$ and hence system-privacy is guaranteed.

\subsection{For arbitrary $N$, $K$, $T$ and $E$ ($E<T$)} \label{sec:general_scheme}
Denote $J =  \frac{N^K-T^K}{N-T}$, and suppose each file comprises  $L = K N^K - E J = K  N^K - E \frac{N^K-T^K}{N-T}$ symbols from a large enough finite field. 
The user downloads $K$ rounds, with $N J$ symbols per round. The database generates $KE J$ uniformly random symbols, denoted by $S_{[1:E J]}^{(r)}$ where $r = [1:K]$.

Divide $[1:L]$ and $[1:EJ]$ into $K$ disjoint sets in the following way, 
\begin{align}
[1:L] & = & \underbrace{\calW_1}_\text{size $N^K \! \! - \! \! EN^{K-1}$} & \cup & \underbrace{\calW_2}_\text{size $N^K \! \! - \! \! ETN^{K-2}$} & \cup \dots & \cup & \underbrace{\calW_{K-1}}_\text{size $N^K \! \! - \! \! ET^{K-2}N$} & \cup & \underbrace{\calW_{K}}_\text{size $N^K \! \! - \! \! ET^{K-1}$} \\
[1:EJ] & = & \underbrace{\calS_1}_\text{size $ EN^{K-1}$} & \cup & \underbrace{\calS_2}_\text{size $ ETN^{K-2}$} & \cup \dots & \cup & \underbrace{\calS_{K-1}}_\text{size $ ET^{K-2}N$} & \cup & \underbrace{\calS_K}_\text{size $ ET^{K-1}$} \label{eqn:EJ}
\end{align}
such that $|\calW_i| + |\calS_i| = N^K$.
Therefore, $W_k^{[1:L]} = \{ W_k^{\calW_1}, \dots, W_k^{\calW_K}\}$ and $S_{[1:E J]}^{(r)} = \{ S_{\calS_1}^{(r)}, \dots,  S_{\calS_K}^{(r)} \}$.

Let $\{ \lambda_1, \dots, \lambda_{N^K} \}$ be $N^K$ distinct nonzero elements from $\Fq$. Let $\bfG$ be a $N^K \times N^K$ matrix defined as follows,
\begin{equation}
{
\bfG= 
\begin{bmatrix}
1 & 1 & \dots & 1\\
\lambda_1 & \lambda_2 & \dots & \lambda_{N^K}       \\
\vdots  & \vdots   & \ddots & \vdots   \\
\lambda_1^{N^K \! - \! 1} & \lambda_2^{N^K \! - \! 1} &  \dots & \lambda_{N^K}^{N^K \! - \! 1}   
\end{bmatrix}
}
, 
\end{equation}
it is direct that $\bfG$ is an invertible matrix. In the following, we divide $\bfG$ into $K$ pairs of matrices $\{ \bfG_{\calW_i}^{|\calW_i| \times N^K},  \bfG_{\calS_i}^{|\calS_i| \times N^K} \}$ for $i = [1:K]$, where $\bfG_{\calW_i}^{|\calW_i| \times N^K}$ is composed of the first $|\calW_i| $ rows of $\bfG$ and $ \bfG_{\calS_i}^{|\calS_i| \times N^K}$ is composed of the last $|\calS_i| $ rows of $\bfG$.
It is direct that these $2K$ matrices are generating matrices of MDS codes with corresponding dimensions, and $\bfG = [\bfG_{\calW_i}^{|\calW_i| \times N^K} \;  \bfG_{\calS_i}^{|\calS_i| \times N^K}  ]^{\textrm{T}}$

For each round $r$ and each file index $k$, let $V_{k}^{(r)}$ be the length-$N^K$ vector defined as follows,
\begin{align}
	V_{k}^{(r)} 
	& = W_k^{\calW_{k+r-1 \bmod K}} \bfG_{\calW_{k+r-1 \bmod K}}^{|\calW_{k+r-1 \bmod K}| \times N^K} + S_{\calS_{k+r-1 \bmod K}}^{(r)} \bfG_{\calS_{k+r-1 \bmod K}}^{|\calS_{k+r-1 \bmod K}| \times N^K}  \\
	& = \left[ W_k^{\calW_{k+r-1 \bmod K}}  \; S_{\calS_{k+r-1 \bmod K}}^{(r)} \right] \bfG, \label{eqn:generalschemeV}
\end{align}
therefore, the $K$ index set pairs $(\calW_i, \calS_i )$ is rotated in all $K$ round for each file index $k \in [1:K]$. This is to assure user-privacy.

The user privately generates $K^2$ matrices  $\bfS_{[1:K]}^{(1)},\bfS_{[1:K]}^{(2)}, \dots, \bfS_{[1:K]}^{(K)} \in \Fq^{N^K \times N^K}$ uniformly and independently from all $N^K \times N^K$ invertible matrices over $\Fq$.

Suppose the user wants to retrieve $W_{l}$. For any undesired file index $k \in [1:K] \setminus \{l\}$, there are $\Delta = 2^{K-2}$ distinct subsets of $[1:K]$ which contain $k$ and do not contain $l$, denoted by $\calK_1, \calK_2, \dots, \calK_{\Delta}$.
For $i \in [1:\Delta]$, let $\alpha_i = N(N-T)^{|\calK_i|-1} T^{K-|\calK_i|}$,
choose $\Delta$ matrices $\bfG_1^{\alpha_1 \times \frac{N}{T}\alpha_1}, \dots, \bfG_{\Delta}^{\alpha_{\Delta} \times \frac{N}{T}\alpha_{\Delta}}$ be the generating matrices of the MDS codes with corresponding dimensions.

For each round $r$, apply the scheme in~\cite{sun2016colluding} for $V_{[1:K]}^{(r)}$ as described in~\eqref{eqn:generalschemeV}. For any undesired file index $k \in [1:K] \setminus \{l\}$,
\begin{align}
X_k^{(r)} & = 
\left[
\begin{array}{c:c:c:c}
x_{\calK_1}^{[k], (r)} \;
x_{\calK_1 \cup \{l\}}^{[k], (r)} 
&
x_{\calK_2}^{[k], (r)} \;
x_{\calK_2 \cup \{l\}}^{[k], (r)} 
&
\cdots 
&
x_{\calK_{\Delta}}^{[k], (r)} \;
x_{\calK_{\Delta} \cup \{l\}}^{[k], (r)}
\end{array} \right] \\
& =
V_{k}^{(r)} \bfS_{k}^{(r)} [:, (1:TN^{K-1})]
\begin{bmatrix}
\bfG_1^{\alpha_1 \times \frac{N}{T}\alpha_1} & \mathbf{0} & \cdots & \mathbf{0} \\
\mathbf{0} & \bfG_2^{\alpha_2 \times \frac{N}{T}\alpha_2} & \cdots & \mathbf{0} \\
\vdots & \vdots & \vdots & \vdots \\
\mathbf{0} & \mathbf{0} & \mathbf{0} & \bfG_{\Delta}^{\alpha_{\Delta} \times \frac{N}{T}\alpha_{\Delta}}
\end{bmatrix},
\end{align}
where the length of $x_{\calK_i}^{[k], (r)}$ is $\alpha_i = N(N-T)^{|\calK_i|-1} T^{K-|\calK_i|}$ and the length of $x_{\calK_i \cup \{l\}}^{[k], (r)} $ is $\frac{N-T}{T} \alpha_i$.

For the desired file index $l$, there are $\delta = 2^{K-1}$ distinct subsets of $[1:K]$ which contain $l$, denoted by $\calL_1, \calL_2, \dots, \calL_{\delta}$. Let
\begin{equation}
X_l^{(r)} 
= \left[ x_{\calL_1}^{[l], (r)} \;  x_{\calL_2}^{[l], (r)} \; \cdots \; x_{\calL_{\delta}}^{[l], (r)} \right]
= V_{l}^{(r)} \bfS_{l}^{(r)},
\end{equation}
where the length of $x_{\calL_i}^{[l], (r)} $ is $N(N-T)^{|\calL_i|-1}T^{K-|\calL_i|}$.

For each non-empty set $\calK \in [1:K]$, the queries associated with $\calK$ is generated by 
\begin{equation}
\calQ_{\calK}^{(r)} = \sum_{k \in \calK} x_{\calK}^{(r)}.
\end{equation}
For all $K$ rounds $r \in [1:K]$, distribute the queries for each $\calK$ evenly among the $N$ databases, and the construction of the queries is completed.

\noindent {\bf Decodability, User-privacy, System-privacy, and the Achievable rate}

From~\cite{sun2016colluding}, for each round, the user can cancel the interference of the undesired files hence obtain $V_{l}^{(r)}$ for all $K$ rounds. Furthermore, from~\eqref{eqn:generalschemeV}, the user can solve for a different set of symbols $W_l^{\calW_i}$ each round, hence the user can obtain all the symbols of the desired file  $W_l^{[1:L]} = \{ W_l^{\calW_1}, \dots, W_l^{\calW_K}\}$.

To see why user-privacy is guaranteed, similarly as in~\cite{sun2016colluding}, any $T$ colluding servers observe queries comprised of $TN^{K-1}$ symbols of $X_{k}^{(r)}$ for each round. Denote the index set of $X_{k}^{(r)}$ observed by the colluding servers by $\calI_k$,  we have that for all $k \in [1:K]$,
\begin{equation}
	X_{\calI_k}^{(r)} \sim V_{k}^{(r)} \bfS_{k}^{(r)}[:, (1:TN^{K-1})].
\end{equation}
From~\eqref{eqn:generalschemeV}, $V_{k}^{(r)} $ are constructed from disjoint set of symbols of $W_{k}$ in an iterative way through the $K$ rounds, and  because $\bfS_{k}^{(r)}[:, (1:TN^{K-1})]$ are independently and identically distributed, user-privacy is guaranteed since the colluding databases observe symbols constructed from all $W_k$'s through the same random mapping.

System-privacy is guaranteed because from~\eqref{eqn:EJ} and~\eqref{eqn:generalschemeV}, for each round the $EJ$ queries and answers observed by the eavesdropper is constructed by adding independent linear combinations of $EJ$ independent uniform symbols$S_{[1:E J]}^{(r)}$. Therefore, the eavesdropper can obtain no information regarding the database $W_{[1:K]}$.

The rate achieved by the scheme is 
\begin{equation}
R = \frac{L}{KNJ} = \frac{K  N^K - E \frac{N^K-T^K}{N-T}}{KN \frac{N^K-T^K}{N-T}} = \frac{1-\frac{T}{N}}{1 - (\frac{T}{N})^K} - \frac{E}{KN} = \underline{R}_{\textrm{T-EPIR}}.
\end{equation}

The secrecy rate achieved is
\begin{equation}
\rho = \frac{KEJ}{L} = \frac{ KE \frac{N^K-T^K}{N-T}  }{  K  N^K - E \frac{N^K-T^K}{N-T}} = \frac{ \frac{E}{N} \left( 1 - (\frac{T}{N})^K \right)}{ 1 - \frac{T}{N} -\frac{E}{KN} \left( 1 - (\frac{T}{N})^K \right) }.
\end{equation}

\bibliographystyle{IEEEtran}
\bibliography{IEEEabrv,PIR}

\begin{thebibliography}{10}
\providecommand{\url}[1]{#1}
\csname url@samestyle\endcsname
\providecommand{\newblock}{\relax}
\providecommand{\bibinfo}[2]{#2}
\providecommand{\BIBentrySTDinterwordspacing}{\spaceskip=0pt\relax}
\providecommand{\BIBentryALTinterwordstretchfactor}{4}
\providecommand{\BIBentryALTinterwordspacing}{\spaceskip=\fontdimen2\font plus
\BIBentryALTinterwordstretchfactor\fontdimen3\font minus
  \fontdimen4\font\relax}
\providecommand{\BIBforeignlanguage}[2]{{%
\expandafter\ifx\csname l@#1\endcsname\relax
\typeout{** WARNING: IEEEtran.bst: No hyphenation pattern has been}%
\typeout{** loaded for the language `#1'. Using the pattern for}%
\typeout{** the default language instead.}%
\else
\language=\csname l@#1\endcsname
\fi
#2}}
\providecommand{\BIBdecl}{\relax}
\BIBdecl

\bibitem{wang2017secure}
Q.~Wang and M.~Skoglund, ``Secure symmetric private information retrieval from
  colluding databases with adversaries,'' \emph{arXiv preprint
  arXiv:1707.02152}, 2017.

\bibitem{chor1995private}
B.~Chor, O.~Goldreich, E.~Kushilevitz, and M.~Sudan, ``Private information
  retrieval,'' in \emph{IEEE Annual Symposium on Foundations of Computer
  Science}, 1995, pp. 41--50.

\bibitem{chor1998private}
B.~Chor, E.~Kushilevitz, O.~Goldreich, and M.~Sudan, ``Private information
  retrieval,'' \emph{Journal of the ACM (JACM)}, 1998.

\bibitem{gertner1998protecting}
Y.~Gertner, Y.~Ishai, E.~Kushilevitz, and T.~Malkin, ``Protecting data privacy
  in private information retrieval schemes,'' in \emph{Proceedings of the
  thirtieth annual ACM symposium on Theory of computing}, 1998.

\bibitem{sun2017capacity}
H.~Sun and S.~A. Jafar, ``The capacity of private information retrieval,''
  \emph{IEEE Transactions on Information Theory}, 2017.

\bibitem{sun2016colluding}
------, ``The capacity of robust private information retrieval with colluding
  databases,'' \emph{arXiv preprint arXiv:1605.00635}, 2016.

\bibitem{sun2016SPIR}
------, ``The capacity of symmetric private information retrieval,''
  \emph{arXiv preprint arXiv:1606.08828}, 2016.

\bibitem{banawan2016capacity}
K.~Banawan and S.~Ulukus, ``The capacity of private information retrieval from
  coded databases,'' \emph{arXiv preprint arXiv:1609.08138}, 2016.

\bibitem{banawan2017multi}
------, ``Multi-message private information retrieval: Capacity results and
  near-optimal schemes,'' \emph{arXiv preprint arXiv:1702.01739}, 2017.

\bibitem{banawan2017capacity}
------, ``The capacity of private information retrieval from byzantine and
  colluding databases,'' \emph{arXiv preprint arXiv:1706.01442}, 2017.

\bibitem{wang2016symmetric}
Q.~Wang and M.~Skoglund, ``Symmetric private information retrieval for {MDS}
  coded distributed storage,'' \emph{arXiv preprint arXiv:1610.04530}, 2016.

\bibitem{wang2017linear}
------, ``Linear symmetric private information retrieval for mds coded
  distributed storage with colluding servers,'' \emph{arXiv preprint
  arXiv:1708.05673}, 2017.

\bibitem{shah2014one}
N.~B. Shah, K.~Rashmi, and K.~Ramchandran, ``One extra bit of download ensures
  perfectly private information retrieval,'' in \emph{Proc. IEEE Int. Symp.
  Information Theory}, 2014, pp. 856--860.

\bibitem{fazeli2015pir}
A.~Fazeli, A.~Vardy, and E.~Yaakobi, ``{PIR} with low storage overhead: coding
  instead of replication,'' \emph{arXiv preprint arXiv:1505.06241}, 2015.

\bibitem{chan2015private}
T.~H. Chan, S.-W. Ho, and H.~Yamamoto, ``Private information retrieval for
  coded storage,'' in \emph{Proc. IEEE Int. Symp. Information Theory}, 2015,
  pp. 2842--2846.

\bibitem{tajeddine2016private}
R.~Tajeddine and S.~E. Rouayheb, ``Private information retrieval from {MDS}
  coded data in distributed storage systems,'' in \emph{Proc. IEEE Int. Symp.
  Information Theory}, 2016.

\bibitem{freij2016private}
R.~Freij-Hollanti, O.~Gnilke, C.~Hollanti, and D.~Karpuk, ``Private information
  retrieval from coded databases with colluding servers,'' \emph{arXiv preprint
  arXiv:1611.02062}, 2016.

\bibitem{cover2012elements}
T.~M. Cover and J.~A. Thomas, \emph{Elements of {i}nformation {t}heory}.\hskip
  1em plus 0.5em minus 0.4em\relax John Wiley \& Sons, 2012.

\end{thebibliography}

\end{document}